\documentclass[twocolumn]{aastex63}
\usepackage{booktabs}
\usepackage[caption=false]{subfig}

\accepted{December 21, 2020}
\submitjournal{AJ}

\newcommand{\IAC}{Instituto de Astrof\'isica de Canarias, E-38205 La Laguna, Tenerife, Spain}
\newcommand{\ULL}{Universidad de La Laguna, Dpto. Astrof\'isica, E-38206 La Laguna, Tenerife, Spain}

\newcommand{\UPitt}{Department of Physics and Astronomy and Pittsburgh Particle Physics, Astrophysics and Cosmology Center (PITT PACC), University of Pittsburgh, Pittsburgh, PA 15260, USA}

\begin{document}

\title{Analysis of Previously Classified White Dwarf-Main Sequence Binaries Using Data from the APOGEE Survey}

\correspondingauthor{Kyle A. Corcoran}
\email{kac8aj@virginia.edu}

\author[0000-0002-2764-7248]{Kyle A. Corcoran}
\affiliation{Department of Astronomy, University of Virginia, Charlottesville, VA 22904, USA}
\author[0000-0002-7871-085X]{Hannah M. Lewis}
\affiliation{Department of Astronomy, University of Virginia, Charlottesville, VA 22904, USA}
\author[0000-0001-5261-4336]{Borja Anguiano}
\affiliation{Department of Astronomy, University of Virginia, Charlottesville, VA 22904, USA}
\author[0000-0003-2025-3147]{Steven R. Majewski}
\affiliation{Department of Astronomy, University of Virginia, Charlottesville, VA 22904, USA}
\author[0000-0002-5365-1267]{Marina Kounkel}
\affiliation{Department of Physics and Astronomy, Western Washington University, Bellingham, WA 98225, USA}
\author[0000-0002-9448-6261]{Devin J McDonald}
\affiliation{Department of Astronomy, University of Virginia, Charlottesville, VA 22904, USA}
\author[0000-0002-3481-9052]{Keivan G. Stassun}
\affiliation{Department of Physics and Astronomy, Vanderbilt University, Nashville, TN 37235, USA}
\author{Katia Cunha}
\affiliation{Steward Observatory, Department of Astronomy, University of Arizona, Tucson, AZ 85721, USA}
\affiliation{Observat{\'o}rio Nacional, Rio de Janeiro, Brasil}
\author{Verne Smith}
\affiliation{NSF OIR Lab, Tucson, AZ 85719, USA}
\author{Carlos Allende Prieto}
\affiliation{\IAC}
\affiliation{\ULL}
\author[0000-0003-3494-343X]{Carles Badenes}
\affil{\UPitt}
\author{Nathan De Lee}
\affiliation{Department of Physics, Geology, and Engineering Technology, Northern Kentucky University, Highland Heights, KY 41099}
\affiliation{Department of Physics and Astronomy, Vanderbilt University, Nashville, TN 37235, USA}
\author{Christine N. Mazzola}
\affil{\UPitt}
\author{Pen{\'e}lope Longa-Pe{\~n}a}
\affiliation{Centro de Astronom{\'i}a, Universidad de Antofagasta, Antofagasta 1270300, Chile}
\author[0000-0002-1379-4204]{Alexandre Roman-Lopes}
\affiliation{Departamento de Astronom{\'i}a, Universidad de La Serena, 1700000 La Serena, Chile}











\begin{abstract}
We present analyses of near-infrared, spectroscopic data from the Apache Point Observatory Galactic Evolution Experiment (APOGEE) survey for 45 previously confirmed or candidate white dwarf-main sequence (WDMS) binaries identified by the optical SDSS and LAMOST surveys.  Among these 45 systems, we classify three as having red giant primaries in the LAMOST sample and fourteen to be young stellar object contaminants in the photometrically identified SDSS sample. From among the subsample of 28 systems that we confirm to have MS primaries, we derive and place limits on orbital periods and velocity amplitudes for fourteen.  Seven systems have significant velocity variations that warrant a post-common-envelope (PCE) binary classification -- four of which are newly classified, three of which are newly confirmed, and five for which we can derive full orbital parameters.  If confirmed, one of these newly discovered systems (2M14544500+4626456) will have the second longest orbital period reported for a typical, compact PCE WDMS binary ($P=15.1$ days).  In addition to the seven above, we also recover and characterize with APOGEE data the well known PCE WDMS systems EG UMa and HZ 9.  We also investigate the overall metallicity distribution of the WDMS sample, which is a parameter space not often explored for these systems.  Of note, we find one system (2M14244053+4929580) to be extremely metal-poor (${\rm [Fe/H]}=-1.42$) relative to the rest of the near-solar sample.  Additionally, the PCE systems in our sample are found to be, on average, higher in metallicity than their wide-binary counterparts, though we caution that with this small number of systems, the sample may not be representative of the overall distribution of WDMS systems.
\end{abstract}



\section{Introduction} \label{sec:intro}
White dwarf stars are the endpoint of stellar evolution for almost every main sequence star. Because most
main sequence stars exist in binary systems \citep{Duquennoy1991,Raghavan2010,Yuan2015}, it is common for the more massive star to evolve into a white dwarf, leading to a detached white dwarf-main sequence (WDMS) binary.  The way in which the more massive star evolves is dictated by its separation from the progenitor binary, but there are two main formation scenarios followed by these systems.  For a majority ($\sim$75\%) the separation is sufficient for the stars to evolve independently of one another, resulting in a wide binary \citep{Willems2004}.  The remaining fraction ($\sim$25\%) of systems can undergo a stage of common envelope (CE) evolution, which causes the orbit to shrink significantly.  Orbital energy can then be deposited into the envelope, which is ejected from the system, leaving behind a close, post-CE (PCE) binary \citep{Webbink2008}.  

The two populations of WDMS binaries also yield a bimodal distribution in their orbital periods.  This was shown in population synthesis studies  \citep[e.g.,][]{Willems2004,Camacho2014,Cojocaru2017}, which predict that wide WDMS binaries should have orbital periods of $P\!>\!100$ days. 
For instance, \citet{Farihi2010}, using high-resolution imaging of 90 white dwarfs with known or suspected low-mass stellar and substellar companions, confirmed observationally that these systems exhibit a bimodal distribution in projected separation; further, they predict that all spatially unresolved, low-mass stellar and substellar companions in their survey will be found to be in short-period orbits.  Indeed, observations of close PCE WDMS binaries show a distribution peaking at $\sim$8 hours \citep{Miszalski2009,Nebot2011}.  It is worth noting, however, that five self-lensing PCE WDMS binaries with early-type companions and larger than typical separations have been found, and four of these have orbital periods of $P>400$ days \citep{Kawahara2018,Masuda2019}.  While still PCE WDMS binaries, these may represent a population with a distinct formation pathway from the typical, more-compact PCE systems.

Apart from the normal migration from intermediate to short periods occurring in the course of normal CE evolution,
PCE systems can evolve to even shorter orbital periods through angular momentum loss due to magnetic braking and/or gravitational wave emission.  As is outlined in \citet{Ren2018} and references therein, it is possible for the PCE system to undergo a second CE stage, eventually producing double-degenerate WDs, cataclysmic variables, or super-soft X-ray sources. 
However, a complete understanding of the variety of these late evolutionary pathways must rest on a better foundational picture of the CE phase, which itself is 
still relatively poorly understood \citep[e.g.,][]{Ivanova2013}.
To guide these theoretical studies, 
more well-characterized systems at all phases of CE and PCE evolution are needed to place firm observational constraints. 
However, identifying systems in the CE phase is challenging, and most efforts are pointed at categorizing and characterizing PCE systems, which are then used to infer the parameters (e.g., envelope ejection efficiencies, angular momentum loss, envelope binding energy)
of the more ``hidden'' CE phase \citep[e.g.,][]{Ivanova2013}.
Of the different kinds of PCE systems (e.g., hot subdwarf B stars, extremely low mass white dwarfs, etc.), PCE WDMS are arguably the most common.  Thus, PCE WDMS systems can play a crucial role in the study of CE evolution \citep{Zorotovic2011}, and are fundamental tools for understanding the range of astrophysically interesting endpoints of that evolution --- e.g., from Type Ia supernovae to gravitational wave sources \citep[e.g.,][]{Toloza2019}.

Significant progress has been made in recent years to identify WDMS binaries using large area spectroscopic sky surveys at optical wavelengths, such as the Sloan Digital Sky Survey \citep[SDSS;][]{York2000,Stoughton2002} and the Large sky Area Multi-Object fiber Spectroscopic Telescope (LAMOST) survey \citep{Zhao2012}.  These systems are identified in the optical through a variety of different methods, such as through $\chi^{2}$-fits of WDMS template spectra covering a vast range of temperatures, gravities, and companion spectral types \citep[see][]{Rebassa-Mansergas2010}, through application of a wavelet transform \citep{Chui1992} that efficiently identifies WDMS spectral features \citep[see][]{Ren2014}, and through color-color cuts such as those in \citet{Rebassa-Mansergas2013}.  
Presently, the most up-to-date catalog of spectroscopically-confirmed WDMS systems identified using SDSS, that by \citet{Rebassa-Mansergas2016}, contains 3294 WDMS binaries.
The photometrically selected catalog of WDMS candidates identified using SDSS from \citet{Rebassa-Mansergas2013} contains 3419 systems.
Meanwhile, \citet{Ren2018} created
a catalog with an additional 876 WDMS binaries identified using LAMOST spectra, 793 of which are claimed to be genuine WDMS systems. 
From these and other works (such as the White Dwarf Binary Pathways Survey --- \citealt{Parsons2016,RebassaMansergas2017}), the total number of systems classified as PCE WDMS binaries stands at $\sim$300 systems, while only $\sim$120 of these have derived orbital periods. 

Fortunately, that number can be increased through a serendipitous channel.
The Apache Point Observatory Galactic Evolution Experiment \citep[APOGEE;][]{Majewski2017} is a high resolution ($R\sim22,500$) infrared (1.5 -- 1.7 $\mu$m) spectroscopic survey that primarily targets red giant stars to study stellar populations across the Milky Way.  However, because of the simple, photometrically-based selection criteria used for APOGEE targeting \citep{Zasowski2013,Zasowski2017}, 
some WDMS systems have received APOGEE observations by chance.  Because of APOGEE's multi-epoch observing strategy, most of these systems 
have high-quality time series radial velocity (RV) information for at least three, and, in some cases, as many as 50 ``visits'' (i.e., epochs).
For systems with 6 or more visits, these data can be used to constrain or derive orbital parameters for these binary systems.

Here we present the 45 systems previously classified as WDMS binaries or candidate WDMS binaries via optical SDSS and LAMOST studies that have also been observed by APOGEE as of November 2019.
Although this is a relatively small subsample drawn from these optical catalogs, the high quality APOGEE data --- which include not only time series RVs, but also the spectroscopically-derived stellar atmospheric parameters and chemistry of the primaries --- not only permit some glimpses into various sources of contamination in these previous, optical WDMS catalogs, but also contribute to the small, but growing census of WDMS systems having detailed characterization of their individual stellar constituents and orbital geometries.
In particular, of the 21 systems that are confirmed here using APOGEE stellar parameters to have MS primaries, twelve have sufficient RV visit information (6+ visits) from APOGEE to allow us to derive, or place limits on, the Keplerian orbital parameters.  The remaining nine systems have more than two visits, which allows us to attempt to place limits on the orbital paramters.
This pilot APOGEE assessment of the specific set of previously known WDMS systems also lends insights into the potential of the greater APOGEE database for not only the identification of previously unknown WDMS and PCE binaries, but to contribute in a major way to the relatively small number of such systems having well characterized system architectures.

\section{The APOGEE Survey} \label{sec:apogee}
The SDSS-III APOGEE \citep{Majewski2017} and SDSS-IV \citep{Blanton2017} APOGEE-2 (Majewski et al., in prep.) surveys are now in their ninth year of observations with the 2.5-m Sloan Telescope \citep{Gunn2006} at Apache Point Observatory in the Northern Hemisphere and their third year of observations with the 2.5-m du Pont Telescope at Las Campanas Observatory in the Southern Hemisphere.  The combined survey databases now encompass more than 2 million spectra of nearly 600,000 distinct stars.  The vast majority of APOGEE targets receive at least three visits, to build up signal-to-noise (S/N) and with the intent that RV variations can be used to identify stars in binary or higher multiplicity systems.  Fainter stars will receive more visits as a means to build up signal sufficient to enable precision chemical abundance analysis on the combined spectra; however, the individual visits, even for faint stars, typically accumulate sufficient flux that good RVs can be derived for more extensive time series exploration.  At the other extreme, a small fraction of APOGEE targets have only one visit for various reasons, but primarily because many of these were obtained as ``bonus'' targets through APOGEE-2 co-observing with the dark time SDSS-IV MaNGA project 
\citep{Bundy2015}.

The APOGEE reduction software \citep{Nidever2015AJ} derives RVs for each star using a two-step process.  First, each visit spectrum is cross-correlated  against a grid of synthetic spectra.  This provides an ``estimated RV'' for each visit which is then used to correct the visit spectra to a common velocity needed for combination to a single, higher S/N spectrum for each source.  The latter can then serve as an intermediate template 
against which the relative velocity of each visit spectrum can be rederived, and the whole process repeated in an iterative fashion to arrive at the best combined spectrum as well as a set of relative RVs. In principle, these well-matching, intermediate cross-correlation template spectra created from the combination of individual visits to a star should yield more precise RVs for that same star than if a synthetic template or the spectrum of another star were employed.  Moreover, this method does not require fore-knowledge of the spectral type of the star to obtain high quality RVs.  On the other hand, it is the case that for some stars (typically the fainter ones) this procedure does not improve on results obtained using synthetic cross-correlation templates, and so the reduction pipeline chooses the better result from the two methods at each iteration in the visit combination/relative RV determination stage of the data processing.

Due to the intrinsic resolution of the APOGEE spectrographs and the fact that they are bench-mounted in tightly controlled vacuum and cryogenic environments \citep{Wilson2019PASP}, the median visit RV precision for APOGEE main survey stars is around 100\,m\,s$^{-1}$ \citep{Nidever2015AJ}.  This is better than is usually employed for the study of stellar binaries, and enables the detection of more subtle RV variability induced by more widely separated and/or lower mass companions \citep[e.g.,][]{Troup2016AJ,Price-Whelan2020ApJ}.

After determination of the per-epoch RVs, the APOGEE reduction pipeline collates the spectra from all epochs after shifting them each to the rest frame velocity.  The final, coadded spectra are then run through the APOGEE Spectral Parameters and Chemical Abundances Pipeline \citep[ASPCAP;][]{Garcia-Perez2016AJ,Jonsson2020} to derive exquisite information on the effective temperatures ($T_{\rm eff}$), gravities ($\log{g}$), and metallicities of each APOGEE target \citep[e.g.,][]{Holtzman2015,Holtzman18}.  \citet{Majewski2017} presents many example spectra of stars spanning large ranges in spectral type and metallicity to showcase the quality of APOGEE spectra, and the publicly available APOGEE spectra can be viewed along with their ASPCAP fits on a per-target basis via the SDSS website.\footnote{\url{https://dr16.sdss.org/infrared/spectrum/search}}  The ASPCAP results for the luminous primaries of the WDMS candidates presented here are included in Table \ref{tab:all_targets} (see columns 6-8), which also gives the 2MASS names for the sources (column 1), the survey in which each system was first identified as a WDMS candidate (LAMOST or SDSS, column 2), the Gaia magnitude (column 3), and the 2MASS $H$ magnitude and $(J-K)$ color (columns 4 and 5, respectively).  We use these data to make an initial global assessment of these 45 systems in Section \ref{sec:cleaning}, aided in part by ancillary information provided by previous studies. That ancillary information is summarized in the final column of Table \ref{tab:all_targets}.

\section{Global Assessment of the Previously Identified WDMS Systems}
\label{sec:cleaning}

As is to be expected, we find here some fraction of contaminants among systems previously identified as WDMS candidates, based on optical spectroscopic and photometric surveys. In this section we identify some categories of contamination, made evident by exploration of the observed global properties of the systems in Table 
\ref{tab:all_targets}.

\begin{figure*}
        \centering
        \subfloat{
            \includegraphics[width=0.48\textwidth]{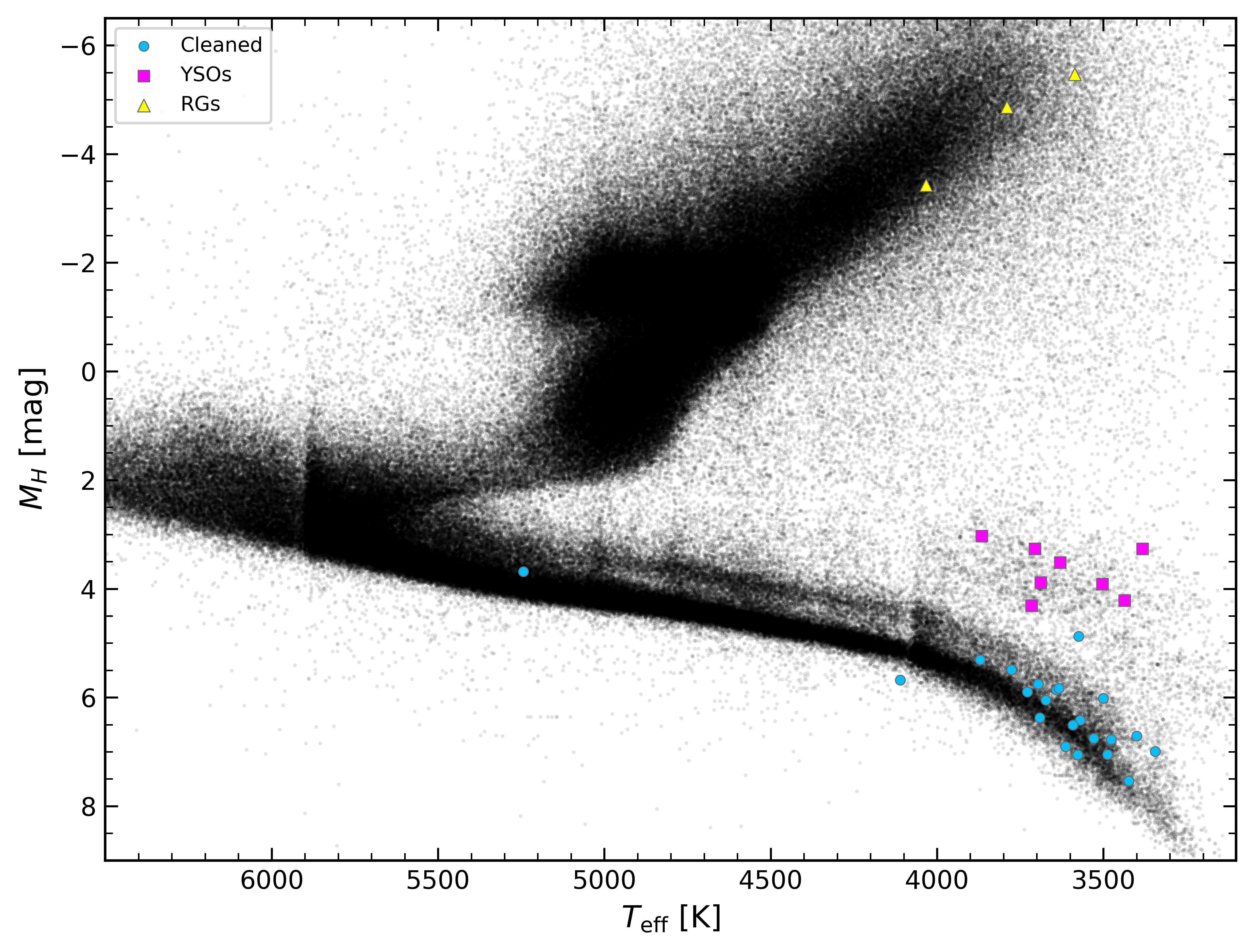}
            \label{fig:CMD}
        }
        \subfloat{
            \includegraphics[width=0.48\textwidth]{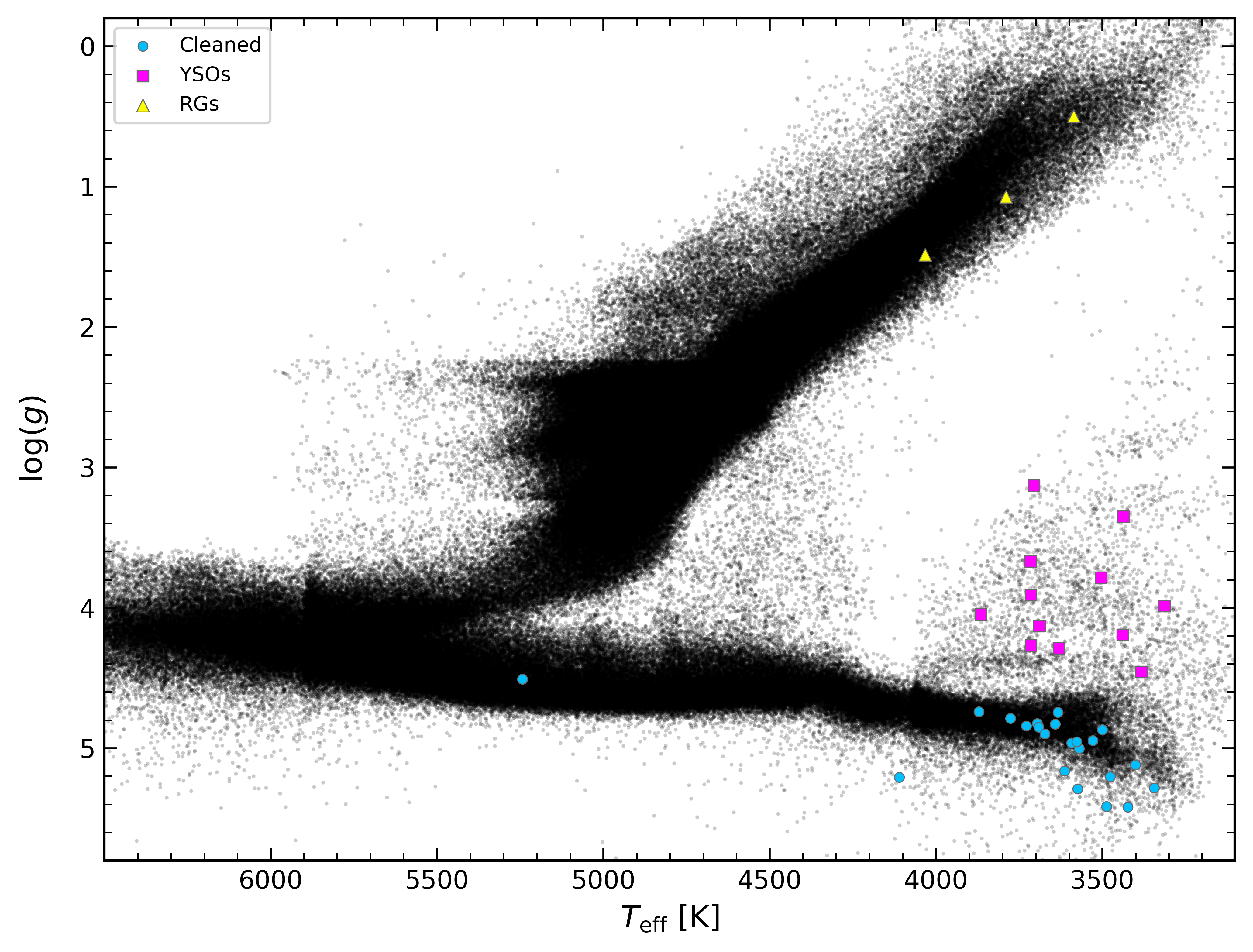}
            \label{fig:Kiel}
        }

        \caption{Confirmed or candidate WDMS systems shown against all of APOGEE.  Note that not all of the 45 systems are shown due to either missing parallaxes (left) or ASPCAP parameters (right).  We highlight in both plots the RG (yellow triangles) and YSO (magenta squares) contaminants as well as the remaining, cleaned sample with MS primaries (blue circles). \textit{Left:} {\it Gaia}-based $H$-band absolute magnitude as a function of the ASPCAP $T_{\rm eff}$. \textit{Right:} Kiel diagram with the ASPCAP $\log(g)$ as a function of ASPCAP $T_{\rm eff}$.}
        \label{fig:CMDs}
\end{figure*}

Figure \ref{fig:CMDs} shows both the {\it Gaia}-based $H$-band absolute magnitude and the ASPCAP-based spectroscopic gravities $\log{g}$ as a function of ASPCAP effective temperature, $T_{\rm eff}$.  The full sample of APOGEE DR16 stars are shown in black and the 45 WDMS candidates as the larger colored dots. The WDMS candidates cluster in three general locations in these observational planes.  The majority of the candidates do indeed lie on the main sequence, as one would hope for WDMS candidates.  Among those systems with MS primaries, the vast majority are at temperatures typical of M dwarfs; only one (2M18454771+4431148) is at a hotter temperature ($T_\mathrm{eff} \sim 5240$ K), a temperature typical of a late G type star.  All of these WDMS candidates on the lower MS have spectroscopically-derived metallicities that are near-solar (within 0.4 dex), consistent with their locations in Figure \ref{fig:CMDs}.

However, one of the systems in the ``lower MS'' group, 
(2M14244053+4929580), actually lies {\it below} the MS at $T_\mathrm{eff} \sim 4110$ K in both panels of Figure \ref{fig:CMDs}.  This ``subdwarf'' location is consistent with the ASPCAP derived metallicity for this star ([Fe/H] = $-1.42$), which shows it to be a rare (see Figure \ref{fig:MDF}, below), metal-poor WDMS candidate.
This metallicity is near the mean value of the Galactic halo \citep{Beers2005}, suggesting that this WDMS system belongs to this population.
Nevertheless, surprisingly, an analysis of the kinematics of this star (Sec. \ref{sec:MDF}) 
shows this binary to actually have a disk-like orbit.

Yet another star in our sample, 2M05303840$-$0525436, has a derived ASPCAP metallicity that is relatively low ([Fe/H] = $-0.8$), but in this case we believe that this metallicity is likely incorrect, because this star is likely a pre-main sequence, young stellar object (YSO; see below). In addition, the spectrum of this star has been flagged by ASPCAP as having potential problems due to cross-talk from a bright neighboring spectrum, as well as landing in parts of the APOGEE detectors that have been identified to have problems with persistence   \citep[see][for an explanation]{Wilson2019PASP}. 
We discuss the subdwarf 2M14244053+4929580 and the overall metallicity distribution function of the cleaned WDMS sample in more detail in Section \ref{sec:MDF}.

Three other stars in Table \ref{tab:all_targets} (2M01090044+5203369, 2M19202987+4000013, and 2M22145972$-$0820200) have both absolute magnitudes and spectroscopically-derived gravities indicating that the primaries are, in fact, on the red giant branch.  LAMOST had classified 2M01090044+5203369 and 2M22145972$-$0820200 as confirmed WDMS binaries, while 2M19202987+4000013 was reported as a candidate WDMS system.  Recently, \citet{Frasca16}, as part of their analysis of the stellar parameters of targets in the {\it Kepler} field, also identified the system 2M19202987+4000013 to have a red giant primary. These three systems, while not WDMS, are still potentially interesting as potential symbiotic star candidates \citep[e.g.,][]{Lewis2020}.

A third group of systems
that can be seen in Figure \ref{fig:CMDs} are those with low $T_\mathrm{eff}$, 
but both higher luminosities and lower surface gravities as compared to normal MS stars of the same temperature. The properties of these systems are consistent with young stellar objects (YSOs). All twelve of these stars are located either in the youngest (1--3 Myr) regions of the Orion Complex \citep{kounkel2018a}, or in NGC 2264, which is another $\sim$3 Myr massive cluster \citep{dahm2008}. Moreover, all twelve of these systems are designated as YSOs in the SIMBAD database \citep{Wenger2000}. 
Two additional systems in Table \ref{tab:all_targets} that do not have derived ASPCAP parameters (due to insufficient S/N or other problems) and therefore do not appear in Figure \ref{fig:CMDs}, 2M05343005$-$0449506 and 2M06402564+0959597, are also identified in SIMBAD to be YSOs.
Given their extreme youth, it is very unlikely that any of these systems have white dwarf companions.

All fourteen of the YSO contaminants were found in the catalog derived from SDSS photometric data, for which the WDMS candidates were selected on the basis of several color--color selections.
However, none of these stars are among those SDSS candidates having previous spectroscopic confirmation by \citet{Rebassa-Mansergas2016}.
Clearly the original optical photometric selection used to select WDMS candidates is also sensitive to YSOs, as evidenced by their prominence in our sample. The main signature of accreting young stars (i.e., Classical T Tauri stars) is strong H$\alpha$ emission, with a 10\% width of the line in excess of 200 km s$^{-1}$, and with equivalent widths that can reach as high as 200 \AA. Weak-Lined T Tauri stars (WTTS, i.e., YSOs that have already stopped accreting and likely have depleted their protoplanetary disks) also can have an H$\alpha$ equivalent width as high as 40 \AA\ in late M stars \citep[e.g.,][]{white2003}.  Apparently such strong Balmer emission, particularly H$\alpha$, confounds 
the various automated WDMS candidate finding algorithms previously employed.

Meanwhile, at shorter wavelengths, an excess of flux in YSOs
can also be a signature of magnetospheric accretion \citep[e.g.,][]{calvet1998a,ingleby2013}, produced by the accretion shocks heating up small spots on the photosphere to temperatures as high as 90,000 K immediately after the shock. Even WTTS systems have NUV luminosities three times higher than what is observed
for MS stars of the same spectral type.  Given their propensity for UV emission, it is understandable how photometric
selection criteria that search for WDMS systems might also recover some YSOs.  Ultimately, it seems that a check of other WDMS candidate properties, such as detection of Li I absorption, placement on the HR diagram, and/or a Galactic position consistent with nearby star forming regions, is needed to identify YSO contaminants from these optically-based WDMS candidate
catalogs. It is also worth noting that while WDMS are multiple systems by definition, comparable photometric or spectroscopic signals could originate from single YSOs without any binary companions.

That about a third of our sample of WDMS candidates turn out to be YSOs may seem surprising, but this should not be interpreted as implying that the SDSS-based (or even the LAMOST-based) WDMS catalog is similarly fractionally contaminated by YSOs. Both the SDSS-based WDMS catalog as well as the APOGEE survey have numerous strong selection biases that complicate interpretation of contamination fractions.  For example, the SDSS survey by and large avoided observations of the Galactic midplane, where YSOs are most concentrated.  On the other hand, APOGEE observations are highly biased {\it towards} the Galactic plane, and, to further amplify that bias toward finding YSOs, the APOGEE survey included a specific focus  
on star forming regions as part of its targeting, in particular through several APOGEE Ancillary Science projects \citep{Zasowski2013,Zasowski2017,cottaar2014,cottle2018}. These competing biases make it difficult to interpret the 27\% YSO contamination of our initial sample, except as a strong alert to this false positive class.

A similar comment may be made regarding the fraction of misidentified giant stars among the WDMS candidates: The $\sim$6\% fraction of giant star systems in our sample, which all come from the LAMOST-based WDMS search, may be inflated by the strong focus on giant stars in APOGEE targeting.  Again, the primary relevant conclusion to be drawn is that the prior WDMS candidate catalogs contain some contamination by misidentified systems with red giant primaries.

\begin{table*}
    \scriptsize
    \centering
    
    \caption{General information on the 45 WDMS systems/candidates observed by APOGEE.}

    \begin{tabular}{@{} lrrrrrrrrr @{}}
    \toprule
    
    \multicolumn{1}{c}{APOGEE ID} & 
    \multicolumn{1}{c}{Source Survey} &
    \multicolumn{1}{c}{$G^{a}$} &
    \multicolumn{1}{c}{$H^{b}$} &
    \multicolumn{1}{c}{$\left(J-K\right)^{b}$} &
    \multicolumn{1}{c}{$T_{\rm eff}\!^{b}$} &
    \multicolumn{1}{c}{$\log \left(g\right)^{b}$} &
    \multicolumn{1}{c}{[Fe/H]$^{b}$} &
    \multicolumn{1}{c}{Notes$^{c}$} \\
    
    \multicolumn{1}{c}{} & 
    \multicolumn{1}{c}{} & 
    \multicolumn{1}{c}{\tiny{[mag]}} &
    \multicolumn{1}{c}{\tiny{[mag]}} &
    \multicolumn{1}{c}{\tiny{[mag]}} &
    \multicolumn{1}{c}{\tiny{[K]}} &
    \multicolumn{1}{c}{\tiny{[cgs]}} &
    \multicolumn{1}{c}{} &
    \multicolumn{1}{c}{} \\

    \midrule
    
    2M01090044$+$5203369 & LAMOST & 11.692 &  8.290$\pm$0.018 & 1.098$\pm$0.032 & 4033$\pm$65 &   1.486$\pm$0.047&  0.254$\pm$0.007 & RG \\
    2M01575656$-$0244460 & SDSS$^{d}$ & 15.041 & 12.103$\pm$0.024 & 0.829$\pm$0.035 & 3871$\pm$86 &   4.739$\pm$0.103&  0.114$\pm$0.016 &  \\
    2M03160020$+$0009462 & LAMOST & 15.363 & 12.069$\pm$0.024 & 0.797$\pm$0.037 & 3574$\pm$77 &   5.290$\pm$0.111&  0.017$\pm$0.019 &   \\
    2M03452349$+$2451029 & LAMOST$^{d}$ & 14.476 & 11.201$\pm$0.030 & 0.880$\pm$0.034 & 3776$\pm$64 &   4.789$\pm$0.110&  0.048$\pm$0.011 &  \\
    2M04322373$+$1745026 & LAMOST & 13.362 & 10.161$\pm$0.019 & 0.840$\pm$0.027 & 3344$\pm$64 &   5.283$\pm$0.110&  0.187$\pm$0.017 &  \\
    2M05242983$+$0023460 & SDSS$^{d}$ & 15.836 & 12.016$\pm$0.023 & 0.974$\pm$0.035 & 3439$\pm$67 &   4.193$\pm$0.122& -0.204$\pm$0.020 & TTau \\
    2M05285461$+$0305035 & SDSS$^{d}$ & 14.681 & 11.141$\pm$0.026 & 0.960$\pm$0.038 & 3630$\pm$62 &   4.288$\pm$0.109&  0.123$\pm$0.012 & RSCVn \\
    2M05303840$-$0525436 & SDSS$^{d}$ & 14.818 & 11.398$\pm$0.024 & 0.991$\pm$0.031 & 3706$\pm$84 &   3.130$\pm$0.057& -0.813$\pm$0.030 & Orion V \\
    2M05321483$-$0620547 & SDSS$^{d}$ & 15.038 & 11.778$\pm$0.026 & 0.923$\pm$0.033 & 3503$\pm$76 &   3.785$\pm$0.044& -0.230$\pm$0.023 & Orion V \\
    2M05325045$-$0035422 & SDSS$^{d}$ & 15.736 & 12.057$\pm$0.022 & 0.975$\pm$0.036 & 3383$\pm$66 &   4.456$\pm$0.124& -0.211$\pm$0.021 & YSO \\
    2M05343005$-$0449506 & SDSS$^{d}$ & 14.427 & 11.265$\pm$0.032 & 1.042$\pm$0.031 &  & & & Orion V \\
    2M05355349$-$0123044 & SDSS$^{d}$ & 16.066 & 11.894$\pm$0.031 & 1.217$\pm$0.034 & 3313$\pm$63 &   3.986$\pm$0.044& -0.261$\pm$0.022 & YSO \\
    2M05361475$-$0613169 & SDSS$^{d}$ & 15.106 & 11.737$\pm$0.032 & 0.973$\pm$0.028 & 3689$\pm$73 &   4.130$\pm$0.118& -0.185$\pm$0.017 & Orion V \\
    2M05393524$-$0436145 & SDSS$^{d}$ & 15.867 & 12.148$\pm$0.024 & 1.244$\pm$0.033 & 3437$\pm$70 &   3.350$\pm$0.047& -0.476$\pm$0.024 & YSO \\
    2M06393441$+$0954512 & SDSS$^{d}$ & 15.585 & 12.215$\pm$0.023 & 1.059$\pm$0.033 & 3715$\pm$78 &   3.910$\pm$0.047& -0.214$\pm$0.018 & Orion V \\
    2M06402564$+$0959597 & SDSS$^{d}$ & 17.083 & 13.130$\pm$0.026 & 1.275$\pm$0.042 &  & & & TTau \\
    2M06404600$+$0917582 & SDSS$^{d}$ & 15.999 & 12.701$\pm$0.026 & 0.881$\pm$0.042 & 3714$\pm$77 &   4.267$\pm$0.112& -0.017$\pm$0.017 & TTau \\
    2M06411837$+$0939411 & SDSS$^{d}$ & 16.031 & 12.649$\pm$0.025 & 0.947$\pm$0.040 & 3716$\pm$88 &   3.669$\pm$0.045& -0.204$\pm$0.023 & Orion V \\
    2M06412562$+$0934429 & SDSS$^{d}$ & 15.521 & 12.391$\pm$0.022 & 0.960$\pm$0.036 & 3866$\pm$84 &   4.048$\pm$0.116& -0.204$\pm$0.018 & TTau \\
    2M08094855$+$3221223 & SDSS$^{d}$ & 16.604 & 13.085$\pm$0.020 & 0.922$\pm$0.035 & 3529$\pm$68 &   4.946$\pm$0.113& -0.016$\pm$0.017 &  \\
    2M08424235$+$5128575 & SDSS$^{d}$ & 13.937 & 10.399$\pm$0.027 & 0.885$\pm$0.030 &  & & &  \\
    2M08531787$+$1147595 & SDSS/LAMOST & 14.454 & 11.544$\pm$0.028 & 0.743$\pm$0.029 &  & & &  \\
    2M09463250$+$3903015 & SDSS$^{d}$ & 15.155 & 11.948$\pm$0.022 & 0.859$\pm$0.030 & 3696$\pm$73 &   4.825$\pm$0.107&  0.091$\pm$0.015 &  \\
    2M10243847$+$1624582 & SDSS & 17.822 & 14.454$\pm$0.054 & 0.771$\pm$0.068 & 3400$\pm$78 &   5.119$\pm$0.113&  0.063$\pm$0.024 &  \\
    2M10552625$+$4729228 & SDSS & 16.664 & 13.227$\pm$0.027 & 0.881$\pm$0.042 & 3500$\pm$74 &   4.868$\pm$0.107&  0.169$\pm$0.018 &  \\
    2M11241545$+$4558412 & SDSS$^{d}$ & 14.603 & 11.396$\pm$0.032 & 0.863$\pm$0.030 & 3691$\pm$71 &   4.849$\pm$0.110&  0.025$\pm$0.014 &  \\
    2M11463394$+$0055104 & SDSS$^{d}$/LAMOST$^{d}$ & 16.741 & 13.397$\pm$0.023 & 0.924$\pm$0.042 & 3476$\pm$76 &   5.204$\pm$0.116& -0.045$\pm$0.022 &  \\
    2M12154411$+$5231013 & LAMOST & 12.581 &  9.340$\pm$0.027 & 0.946$\pm$0.030 & 3487$\pm$56 &   5.414$\pm$0.114&  0.028$\pm$0.012 &  \\
    2M12333939$+$1359439 & SDSS$^{d}$/LAMOST$^{d}$ & 16.679 & 13.486$\pm$0.030 & 0.922$\pm$0.036 &  & & &  \\
    2M12423245$-$0646077 & SDSS & 16.065 & 12.968$\pm$0.027 & 0.915$\pm$0.037 & 3728$\pm$78 &   4.842$\pm$0.112& -0.059$\pm$0.017 &  \\
    2M13054173$+$3037005 & SDSS$^{d}$ & 16.237 & 12.761$\pm$0.022 & 0.803$\pm$0.033 & 3424$\pm$72 &   5.420$\pm$0.119& -0.113$\pm$0.022 &  \\
    2M13090450$+$1411351 & SDSS$^{d}$/LAMOST$^{d}$ & 15.308 & 12.229$\pm$0.023 & 0.845$\pm$0.033 &  & & &  \\
    2M13115337$+$1549147 & SDSS$^{d}$ & 16.825 & 13.353$\pm$0.031 & 0.927$\pm$0.048 & 3570$\pm$73 &   5.003$\pm$0.117& -0.142$\pm$0.019 &  \\
    2M13463968$-$0031549 & SDSS$^{d}$ & 16.427 & 13.382$\pm$0.035 & 0.823$\pm$0.046 &  & & &  \\
    2M14244053$+$4929580 & SDSS & 15.694 & 13.542$\pm$0.031 & 0.631$\pm$0.052 & 4111$\pm$111 &   5.208$\pm$0.160& -1.415$\pm$0.038 &  \\
    2M14544500$+$4626456 & LAMOST$^{d}$ & 14.531 & 11.278$\pm$0.032 & 0.925$\pm$0.035 & 3673$\pm$71 &   4.898$\pm$0.106&  0.148$\pm$0.014 &  \\
    2M14551261$+$3810342 & LAMOST & 14.400 & 11.279$\pm$0.022 & 0.802$\pm$0.029 & 3641$\pm$84 &   4.829$\pm$0.122& -0.308$\pm$0.024 &  \\
    2M15041191$+$3658150 & LAMOST & 17.034 & 13.722$\pm$0.036 & 0.873$\pm$0.043 & 3592$\pm$71 &   4.962$\pm$0.117& -0.130$\pm$0.017 &  \\
    2M15104562$+$4048271 & SDSS/LAMOST & 14.671 & 11.374$\pm$0.019 & 0.796$\pm$0.028 & 3577$\pm$78 &   4.954$\pm$0.127& -0.372$\pm$0.023 &  \\
    2M15150334$+$3628203 & SDSS$^{d}$ & 14.941 & 11.652$\pm$0.022 & 0.906$\pm$0.029 & 3634$\pm$65 &   4.744$\pm$0.104&  0.240$\pm$0.012 &  \\
    2M18454771$+$4431148 & LAMOST & 13.742 & 12.075$\pm$0.020 & 0.445$\pm$0.025 & 5243$\pm$131 &   4.509$\pm$0.085&  0.093$\pm$0.012 &  \\
    2M19202987$+$4000013 & LAMOST$^{d}$ & 12.856 &  9.345$\pm$0.021 & 1.084$\pm$0.025 & 3789$\pm$61 &   1.071$\pm$0.046&  0.135$\pm$0.009 & RG \\
    2M19274650$+$3841111 & LAMOST & 12.006 &  7.951$\pm$0.016 & 1.208$\pm$0.038 & 3586$\pm$60 &   0.497$\pm$0.050& -0.326$\pm$0.013 & RG \\
    2M22145972$-$0820200 & SDSS & 17.427 & 14.189$\pm$0.043 & 0.894$\pm$0.069 &  & & &  \\
    2M22200576$-$0418445 & LAMOST & 15.387 & 12.405$\pm$0.023 & 0.792$\pm$0.035 & 3613$\pm$86 &   5.161$\pm$0.122& -0.306$\pm$0.026 &  \\
         
    \bottomrule
    
    \multicolumn{7}{@{}l}{\textbf{\scriptsize{Notes}}} \\
    \multicolumn{7}{@{}l}{\tiny{$^{a}$ From \citet{GaiaCollab2018}}} \\
    \multicolumn{7}{@{}l}{\tiny{$^{b}$ From APOGEE}} \\
    \multicolumn{7}{@{}l}{\tiny{$^{c}$ Descriptions follow the condensed object descriptions from SIMBAD}} \\
    \multicolumn{7}{@{}l}{\tiny{$^{d}$ SDSS photometric candidate or LAMOST spectroscopic candidate}} \\
    
    \end{tabular}
    
    \label{tab:all_targets}
\end{table*}

\section{Analysis of System Architectures}
\label{sec:sytem_architectures}

After removing the YSO and red giant systems from the initial set of 45 previously identified WDMS systems, we are left with 28 remaining as candidate WDMS systems.  Of these, 21 have two or more epochs of RV data, which allow either limits on or solutions to the system orbital parameters.

\subsection{{\it The Joker} Orbital Analysis of the Radial Velocities}
\label{sec:RVs}
Analysis of the multi-epoch RVs 
was performed using \textit{The Joker} \citep{Price-Whelan2017ApJ,Price-Whelan2020ApJ}, a custom Monte Carlo sampler that uses a given set of input RV measurements to produce independent posterior samples in Keplerian orbital parameters.  In particular, the code was designed to excel at fitting orbits for targets with sparse RV data and/or low S/N RV measurements.  Here we provide a brief description of the fitting procedure for clarity, but a more thorough and technical prescription can be found in \citet{Price-Whelan2017ApJ,Price-Whelan2020ApJ}. First, $2^{24}$ samples are drawn from a prior probability density function covering the full Keplerian orbit parameter space, allowing rejection sampling over a dense set of potential solutions.  For systems that have a large number of surviving samples (in this work, 256 samples) that are not unimodal, we recompute the rejection sampling with 512 requested samples
in an attempt to discover groupings of possible solutions in the period distribution and period versus eccentricity diagram that can place limits on the orbital parameters (see Sec.~\ref{sec:WBs} and \ref{sec:25visits} for further details).  If the number of surviving samples is fewer than 256 samples, however, these surviving samples are used to initialize a Markov chain Monte Carlo (MCMC) run.  This procedure typically returns a unimodal set of samples that represent the best fitting solution for a system.

To limit our analysis
to only RVs derived from high-quality spectra, we remove any visit-level APOGEE data that have the following \texttt{STARFLAG}s\footnote{\url{https://www.sdss.org/dr16/algorithms/bitmasks/\#APOGEE_STARFLAG}} set: \texttt{LOW\_SNR} (visit-$\mathrm{S/N}<5$), \texttt{VERY\_BRIGHT\_NEIGHBOR} (indicates that a star with a spectrum adjacent to that of the target star on the spectrograph detector is more than 100 times brighter and therefore a source of potential contaminating flux), \texttt{PERSIST\_HIGH} (evidence that the spectrum crosses detector pixels that show super-persistence), \texttt{PERSIST\_JUMP\_POS}, or \texttt{PERSIST\_JUMP\_NEG} (indicating an obvious and artificial positive or negative decrease, or ``jump", in spectral continuum between two spectrograph detectors --- the one sampling from the 1.585 to 1.644\,$\mu$m and the other from 1.514 to 1.581\,$\mu$m). These flags correspond to bitmask values: 3, 4, 9, 12, 13.

Because the APOGEE visit-level RV uncertainties (\texttt{VRELERR} in the allVisit file) are known to be underestimated \citep[e.g.,][]{Badenes2018ApJ}, 
for the systems surviving the above target flag pruning we apply the expression presented in Brown et al.~(in prep.) 
\begin{equation}
\small{\sigma_\mathrm{RV}^2 = (3.5 (\texttt{VRELERR})^{1.2})^2 + (0.072\,\mathrm{km\,s}^{-1})^2,}
\end{equation}
where $\sigma_\mathrm{RV}$ is the total, inflated visit velocity error for a given visit.  This asymptotes to a $0.072$\,km\,s$^{-1}$ minimum for the visit-level RV uncertainties.

For systems with $\gtrsim$8 visits that are well distributed in orbital phase, \textit{The Joker} generally converges to a single, unimodal period solution.  This is because  \textit{The Joker} fits six orbital elements, i.e., has 6 degrees of freedom.
Of the 21 WDMS binary candidates having at least two APOGEE RV visits, eight have at least eight visits, and thus are good candidates for full Keplerian orbit fitting with \textit{The Joker}, although not all will return satisfactory solutions because of  undetectable orbital amplitudes in the cases of wide binaries. 
The four systems with six or seven RV visits often converge to a single solution, but not universally.
In cases where no single solution is reached (i.e., those with multimodal period solutions), only limits can be placed on the period of the system. This is typical for the nine systems with two to five visits.

In the end, 
we report full orbital solutions for five short period (relative to the APOGEE temporal baseline) systems.  In all five cases, these are the first solutions ever presented for these systems, including RV variations that warrant a PCE classification.  We discuss these five systems in more detail in Sections \ref{sec:2M10243847+1624582}-\ref{sec:2M14544500+4626456}.  For another seven systems that are sampled reasonably to very well by APOGEE (i.e., $\ge6$ RV epochs) no good solution converges because the period of the WDMS binary may be longer than their APOGEE time series data; therefore we report them as ``wide binary'' systems and provide lower limits on the orbital period and the main sequence star velocity amplitude.  These systems are discussed in Section \ref{sec:WBs}.  For an additional nine systems having 2-5 visits, we also provide upper and lower limits to the orbital period (Sec.~\ref{sec:25visits}), and among these are two well known PCE systems and two newly discovered systems that have RV variations warranting a PCE classification.  A few of the WDMS systems observed by APOGEE are left with 1 or 0 useful visits after imposition of the quality cuts described in Section \ref{sec:RVs}, and we briefly mention these in Section \ref{sec:onevisit}.

\subsection{Estimating the Stellar Masses}
\label{sec:stellar_masses}

Along with the orbital parameters provided by {\it The Joker}, we can estimate the primary star masses 
via the empirically-derived \cite{Torres2010}  relation for main sequence stars:
\begin{eqnarray}
    \log{M_\star} & = a_1 + a_2X + a_3X^2 + a_4X^3 + a_5(\log{g})^2 \nonumber \\
    & + a_6(\log{g})^3 + a_7[\mathrm{Fe/H}],
\end{eqnarray}
which has a relatively small scatter ($\sigma_{M_\star} = 0.064 M_\star$) for stars down to $\sim$0.5\,$M_\odot$. Here, $X = \log{T_\mathrm{eff}} - 4.1$ and the coefficients, $a_i$, are given in \cite{Torres2010}.  For the systems with unimodal samples returned by {\it The Joker}, we additionally calculate the minimum masses of the secondary (i.e., the WD companions), $m \sin{i}$ using the calculated primary masses.
In Table \ref{tab:all_orbits}, we report the masses for all the MS stars in our clean sample that have ASPCAP parameters, and we report the minimum masses of the WD stars and derived orbital parameters in systems with well-constrained orbital periods.

\section{Descriptions of Individual Systems}
\label{sec:individual_systems}

\subsection{2M10243847+1624582}\label{sec:2M10243847+1624582}
\begin{figure}[h]
    \centering
    \includegraphics[width=\linewidth]{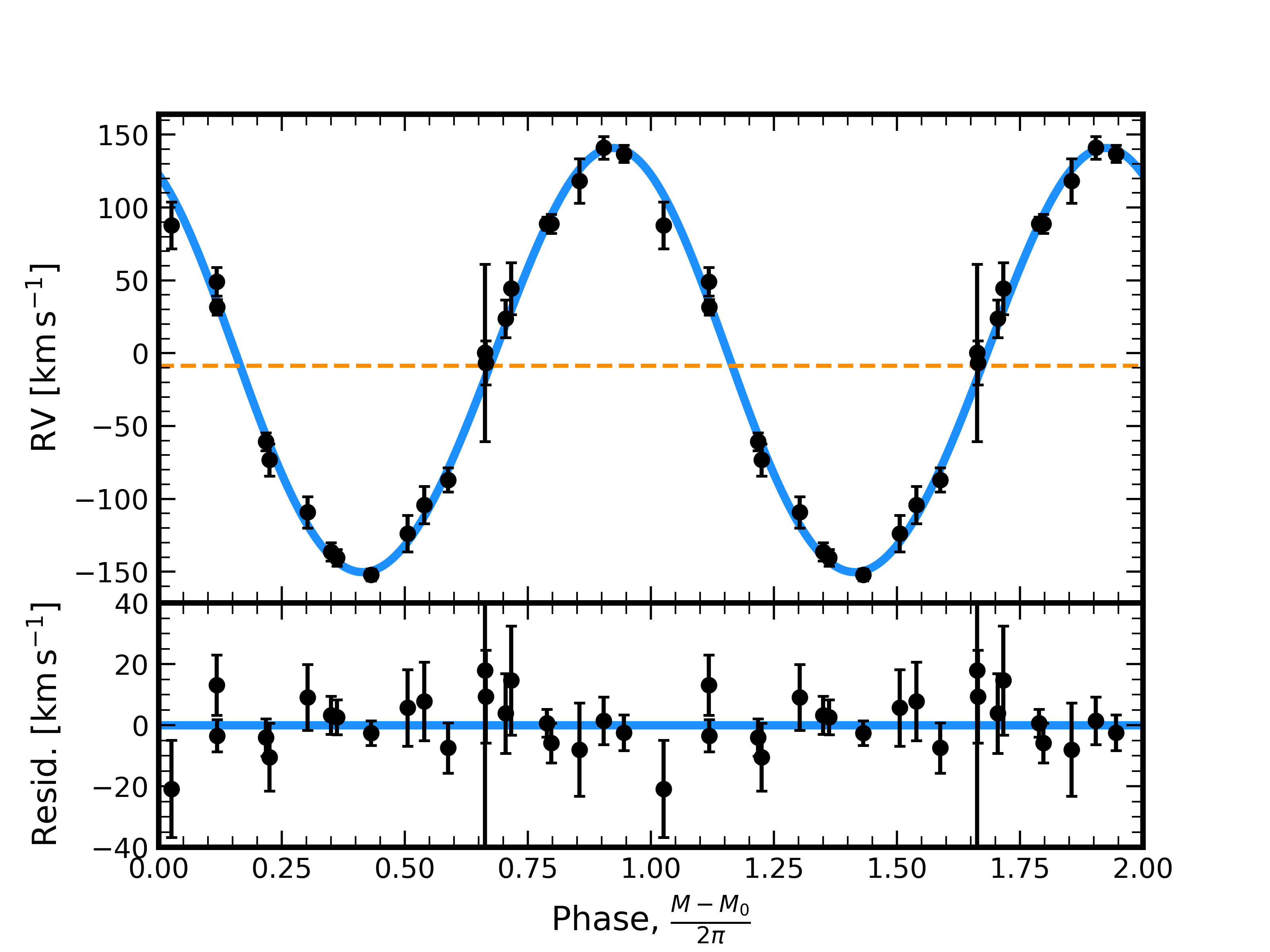}
    \caption{{\it Upper panel:} The best fitting RV curve derived from the orbital solution to the 28 phase-folded APOGEE RVs for the WDMS system 2M10243847+1624582.  Two periods are shown for clarity. {\it Lower panel:} The residuals to the fit shown in the upper panel.}
    \label{fig:2M10243847+1624582}
\end{figure}

High-quality APOGEE observations of the spectroscopically-confirmed WDMS system 2M10243847+1624582 exist for 28 epochs spanning just over a year, from 2015 February to 2016 February.  APOGEE spectroscopic analysis reveals the cool ($T_{\rm eff}$ = 3400 K) M dwarf primary to be approximately solar metallicity ([Fe/H]=$+0.06$). The combined {\it The Joker} plus MCMC run returned a unimodal solution, and this best-fit solution is presented in Figure \ref{fig:2M10243847+1624582}.  With an RV semi-amplitude of 146 km s$^{-1}$ and an orbital period of just over 12 hours, this system is fairly typical for a PCE WDMS system. \citet{Schreiber2010} identified this system as a PCE candidate, and our solution confirms this classification.  The lower limit on the WD mass of $m \sin{i}$ = 0.537 $\pm$0.013 M$_{\odot}$ is consistent with typical masses reported by both \citet{Rebassa-Mansergas2016} for all SDSS systems and \citet{Ren2018} for all LAMOST systems. 
In this case, however, the mass derived from SDSS with a combined atmospheric parameters and cooling track fitting is $M_{WD}=0.830\pm0.063$ $M_{\odot}$, which is well above our derived mass limit, possibly indicating a slightly inclined orbit.

\subsection{2M10552625+4729228}\label{sec:2M10552625+4729228}
\begin{figure}[h]
    \centering
    \includegraphics[width=\linewidth]{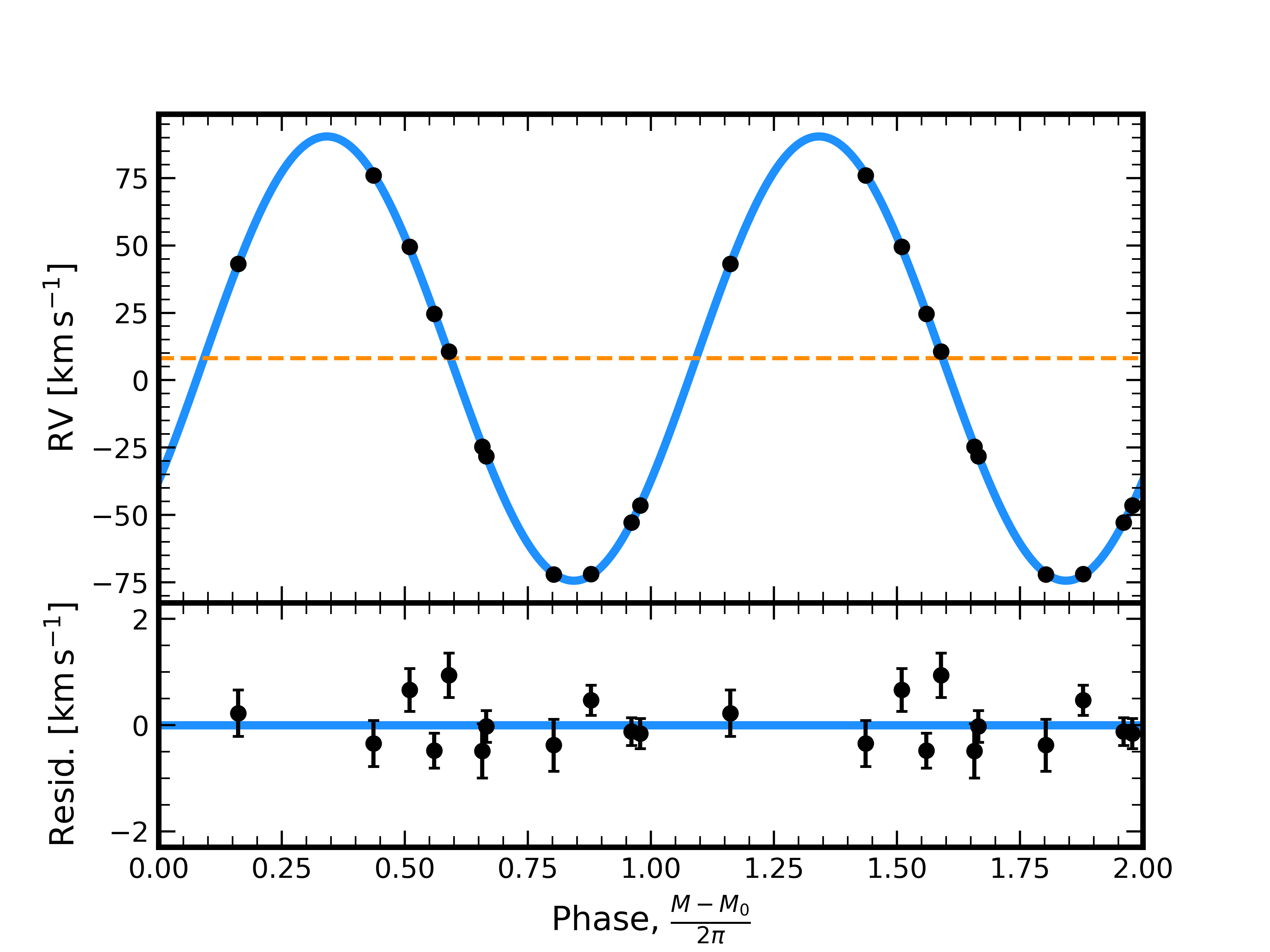}
    \caption{The same as Figure \ref{fig:2M10243847+1624582} but for the eleven APOGEE RV epochs for 2M10552625+4729228.  The error bars are smaller than the plotted points in the upper panel.}
    \label{fig:2M10552625+4729228}
\end{figure}

The WDMS system 2M10552625+4729228 was observed with high S/N eleven times 
over 
277 days from 2017 May to 2018 February.  APOGEE spectroscopic analysis again reveals a cool ($T_{\rm eff}$ = 3500 K) M dwarf primary, this one with a super-solar metallicity ([Fe/H]=+0.17) Our analysis with {\it The Joker} yields a unimodal solution to the RV variations of the M dwarf primary, 
shown in Figure \ref{fig:2M10552625+4729228}. The solution reveals this binary to be another PCE system; with a derived period of a little more than 2 days, though slightly longer than the 8 hour average for PCE WDMS systems, the system is still in a much shorter period orbit than the longest known for PCE binaries.  \citet{Schreiber2010} also identified this system as a PCE candidate, and our solution confirms this classification.   
The lower limit on the WD mass of $m \sin{i}$ =  0.476 $\pm$0.009 M$_{\odot}$ is  similar to that of 2M10243847+1624582, which again could indicate a slightly inclined orbit since it is well below the SDSS mass of $M_{WD}=0.790\pm0.081\,M_{\odot}$.  This value is, however, consistent with typical SDSS and LAMOST WD masses.

\subsection{2M11463394+0055104}\label{sec:2M11463394+0055104}
\begin{figure}[h]
    \centering
    \includegraphics[width=\linewidth]{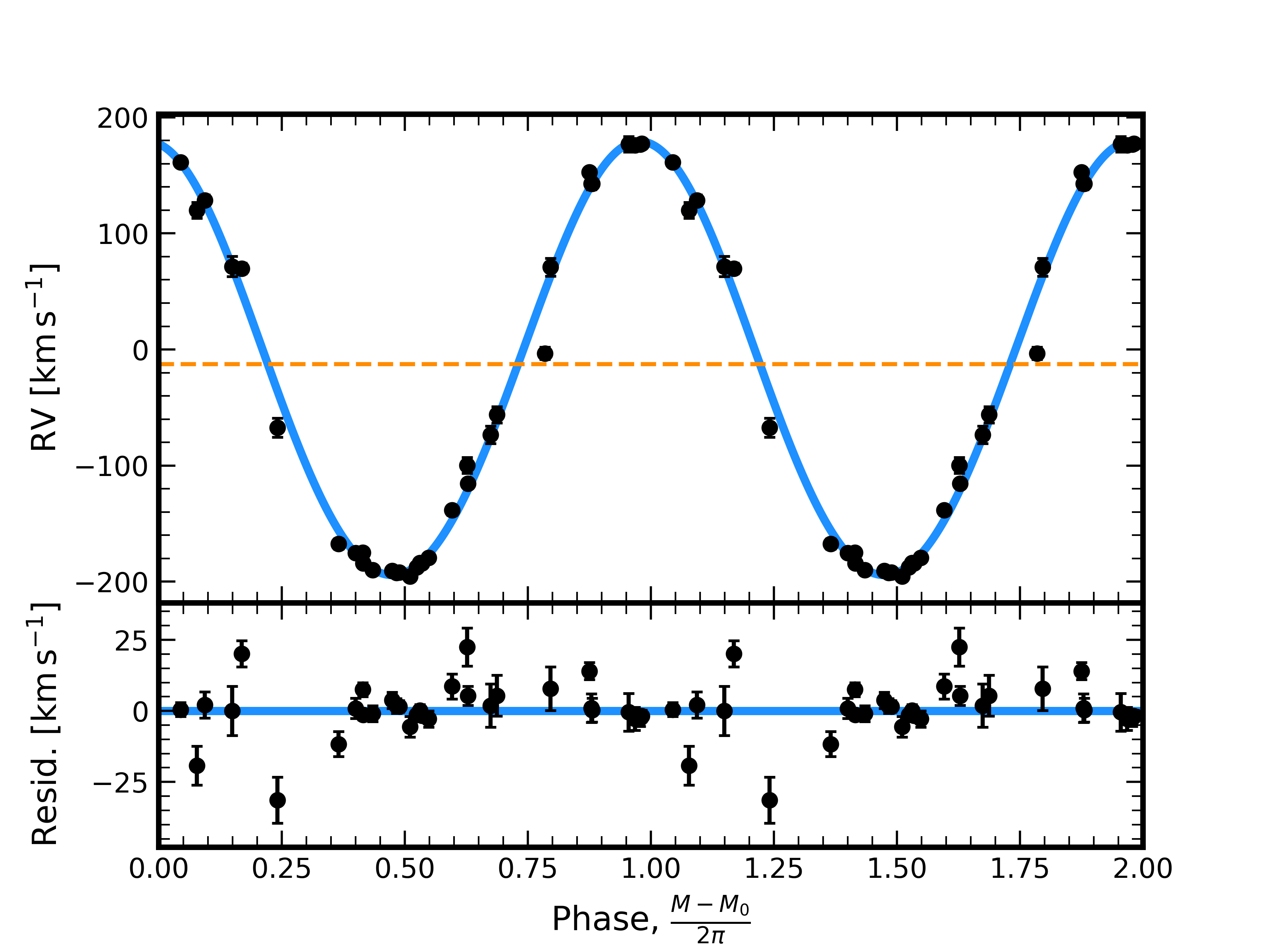}
    \caption{The same as Figure \ref{fig:2M10552625+4729228} but for the 33 APOGEE RV epochs for 2M11463394+0055104. }
    \label{fig:2M11463394+0055104}
\end{figure}

The target 2M11463394+0055104, determined by APOGEE spectroscopic  analysis to be another approximately solar metallicity M dwarf ([Fe/H]=$-0.05$), received 33 high-quality visits spanning just over 4 years from 2012 February to 2016 March.  With such extensive coverage, it is not surprising that a unimodal solution was achieved for this system; the best-fit solution is shown in Figure \ref{fig:2M11463394+0055104}. A roughly 9 hour orbital period places this system just above the typical 8 hour period for PCE WDMS systems.  LAMOST flagged this system as a candidate WDMS binary based on their spectra, and, combined with our derived orbital solution, we confirm this system to be another PCE WDMS binary.  The derived lower limit to the WD mass of $0.717\pm0.014$ $M_\odot$ places this companion near the upper end of the mass distributions reported by both SDSS and LAMOST.  While this is not atypical nor the most massive WD by far, this larger mass is worth pointing out, 
since it is in contrast to the mass limits
of the other systems with unimodal solutions.  

\subsection{2M13054173+3037005}\label{sec:2M13054173+3037005}
This target received 
only six high-quality visits spanning just over a month from 2018 April 24 to 2018 May 31.  To compensate for the smaller number of RV epochs, we decrease the number of degrees of freedom in the solution by setting the eccentricity of all attempted fits to $e=0$ (as we do not have sufficient data to prefer an eccentric orbit over a circular one at this time).
We then performed runs with \textit{The Joker}
using a succession of minimum periods starting with 0.055, 0.1, 0.23, 0.3, 0.64, 0.73, and 1.1 days and a maximum period of 7.0 days; that maximum period was selected because the first four RVs collected looked like they formed a possible full orbital period of that length.  After running \textit{The Joker} on this system multiple times, it was evident that there were multiple solutions that fit the data even though any individual run could return a single solution depending on the specified minimum orbital periods.
With only six data points, the phase coverage and RV spread (${\rm RV_{\rm max}- RV_{\rm min}}=\Delta \rm RV_{max}$, as in Table \ref{tab:all_orbits}) for all solutions in one {\it The Joker} run of a given set lower period limit were inconsistent with one another,
and not all of the runs returned realistic solutions.  
The solution we report was selected because it
contains the most-well-sampled phase coverage over one proposed orbital period while also aligning well with the current RV spread; however, it is important to stress that is only one of many potential solutions to the data. 
This solution shows an orbital period of $P=0.22$ days with the eccentricity fixed to $e=0$, which is a reasonable assumption for such a short orbital period, given that we do not have any reason to believe that such a small orbit would not be circularized.
There were shorter period solutions when \textit{The Joker} was given a shorter minimum period parameter, but the current RV spread appeared far too small to warrant these solutions as they require the $\Delta \rm RV_{max}$ to be $\sim$30--40 $\rm \:km\,s^{-1}$ 
larger than observed.
There are also longer period solutions that one can achieve by imposing a larger minimum orbital period; however, these solutions
generally depended on the observations being poorly distributed in orbital phase, and, in some case, bunch the observations in such a way as to also, 
ironically, result in 
$\Delta \rm RV_{max}$ that are $\sim$30--40 $\rm \:km\,s^{-1}$ larger than observed.
The period and $m\sin{i}$ from our adopted solution are
likely lower limits, 
and clearly additional data are required to fully constrain the orbital parameters for this system.  This system comes from the \citet{Rebassa-Mansergas2013} catalog of photometrically selected WDMS candidates, and we keep this classification and update it to be a PCE candidate given our updated contributions to the RV variation.
\begin{figure}[h]
    \centering
    \includegraphics[width=\linewidth]{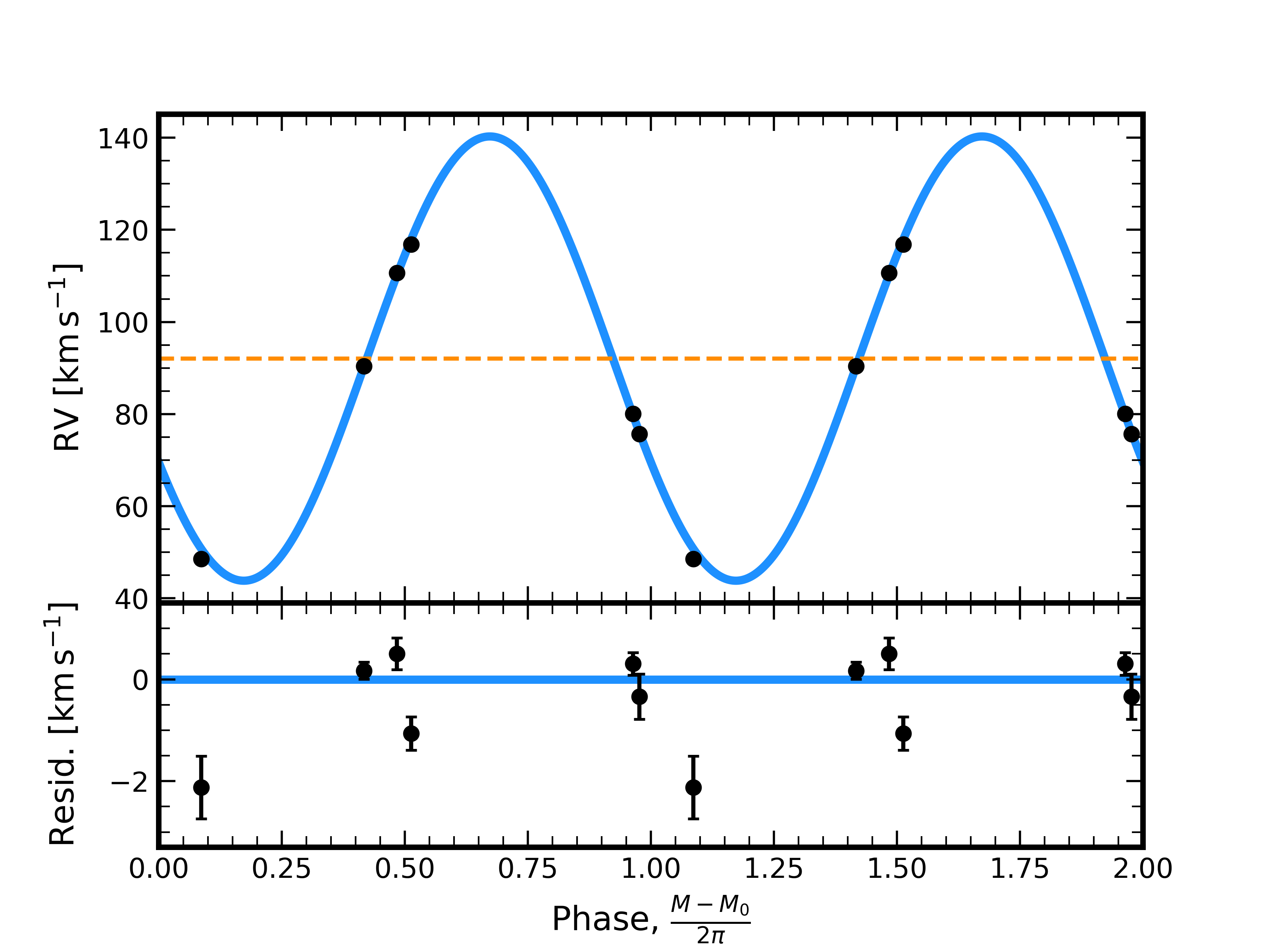}
    \caption{The same as Figure \ref{fig:2M10552625+4729228} but for the six APOGEE RV epochs for 2M13054173+3037005 and assuming a circular orbit.  While the number of epochs is small, their phase coverage is very good {\it for this adopted solution}, while the RV amplitude is very large, which results in a solution that converges to a tight match with the data.}
    
    \label{fig:2M13054173+3037005}
\end{figure}

\subsection{2M14544500+4626456}\label{sec:2M14544500+4626456}
\begin{figure}[h]
    \centering
    \includegraphics[width=\linewidth]{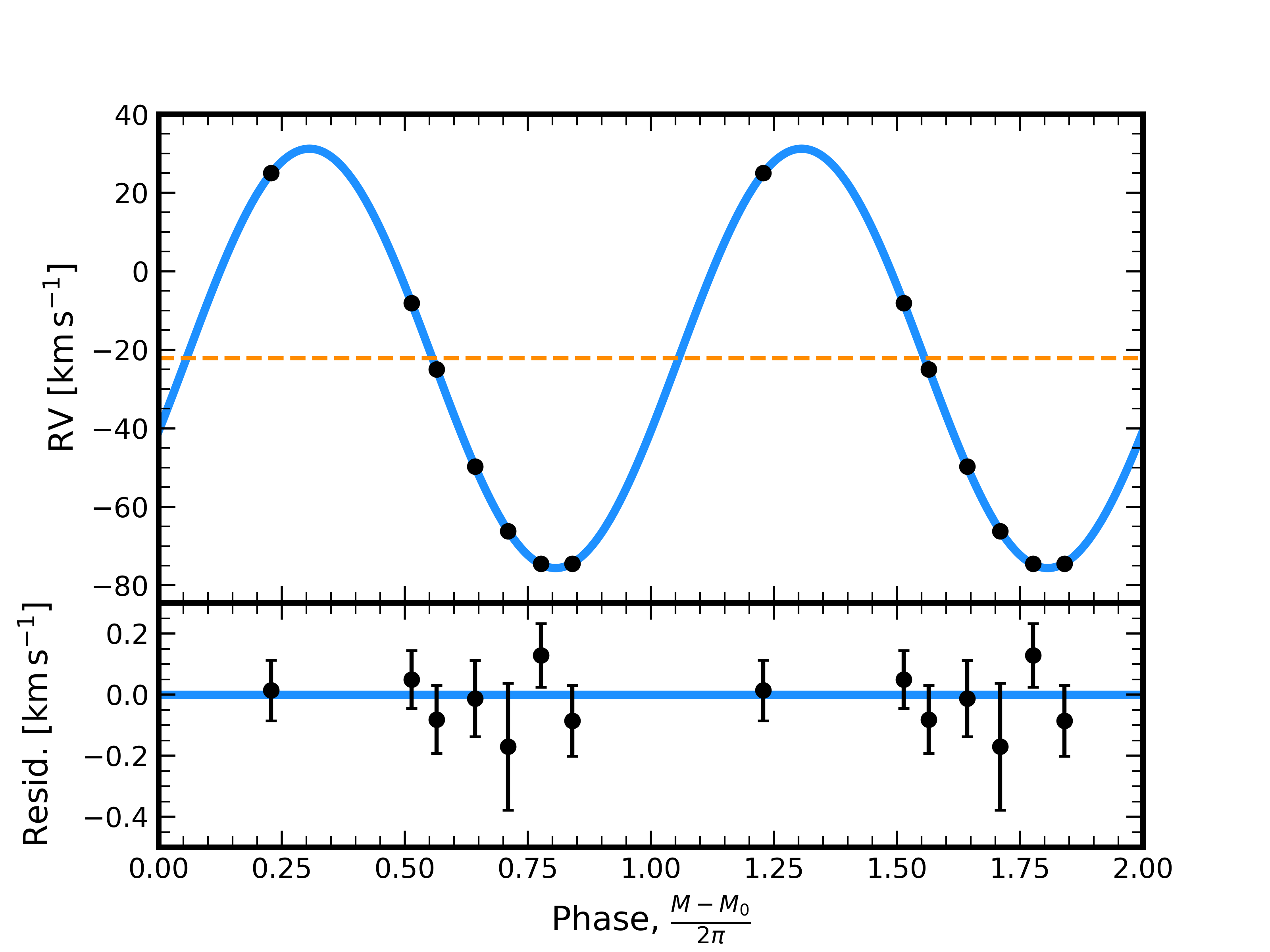}
    \caption{The same as Figure \ref{fig:2M10552625+4729228} but for the seven APOGEE RV epochs for 2M14544500+4626456.  While the number of epochs is relatively small, the combination of very good phase coverage and a large RV amplitude means that a very good solution is possible, under the assumption of a circular orbit. }
    \label{fig:2M14544500+4626456}
\end{figure}

APOGEE observed this target
with seven high-quality APOGEE visits spanning almost four years from 2013 March to 2017 March.  APOGEE spectroscopic analysis again reveals an M dwarf primary with super-solar metallicity ([Fe/H]=+0.15).  \textit{The Joker} returned a unimodal solution for this system, and an MCMC run confirmed that result shown in Figure \ref{fig:2M14544500+4626456}.  Similarly to the situation with 2M13054173+3037005, we fix the eccentricity to be $e=0$ to limit the degrees of freedom as we do not currently have sufficient data to prefer an eccentric orbit at this time.  As with 2M11463394+0055104, our solution yields a minimum WD mass of $m \sin i = 0.693\pm0.001$ $M_\odot$, which places the mass near the upper end of the SDSS and LAMOST distributions.  The orbital period corresponding to this solution is $P=15.10$ days, which would make this system second longest known orbital periods for the typical, compact PCE WDMS systems, behind the five self-lensing systems \citep[$P\sim88-683\;$days;][]{KruseAlgol2014,Kawahara2018,Masuda2019} as well as IK Peg \citep[$P=21.72\;$days;][]{Vennes1998} and just above SDSS J222108.45+002927.7 and SDSS J121130.94$-$024954.4 \citep[$P=9.59\;$days and $P=7.82\;$days, respectively;][]{Rebassa-Mansergas2012}.  With the given number of visits, the phase sampling could be such that this solution is not an accurate model for the system.  Additional RV data will be necessary to adjust or falsify this model in the future; however, we report the best-fit solution we derived as the most likely estimate for the period given the APOGEE data.  As this system was flagged as a candidate WDMS system by LAMOST, this classification will need to be confirmed.  If the WDMS classification and our orbital solution is confirmed, 2M14544500+4626456 will be a useful PCE system for applying the methods in \citet{Rebassa-Mansergas2012} to explore the earlier phases of such systems and the energy budget of CE evolution.

\subsection{Wide Binaries}\label{sec:WBs}
There are seven systems with 6+ high quality epochs that have a small ($\Delta{\rm RV_{\rm max}}<10-20\,{\rm km\,s}^{-1}$) RV spread that may correspond to $K_{\rm MS}<5-10\,{\rm km\,s}^{-1}$.  This low RV spread could be due to a lower inclination angle, which would make detecting shifts in the radial velocity more difficult.  As we cannot place limits on this inclination given the present data, we proceed under the assumption
that these relatively small shifts are a by-product of a system with a large orbital separation and long orbital period.  Indeed, \citet{Willems2004} showed that wide-WDMS binaries should typically have small RV shifts $\left(K_{\rm MS}\sim 1-5\ {\rm km\,s}^{-1}\; {\rm at}\; i=60^{\circ}\right)$ at longer orbital periods $\left(P>100\;{\rm days}\right)$.  For this reason, we classify these seven systems as wide binaries, 
with the understanding that lower inclination angles could mean that we simply cannot detect more massive, unresolved systems. 

\begin{figure*}
        \centering
        \subfloat{
            \includegraphics[width=0.48\textwidth]{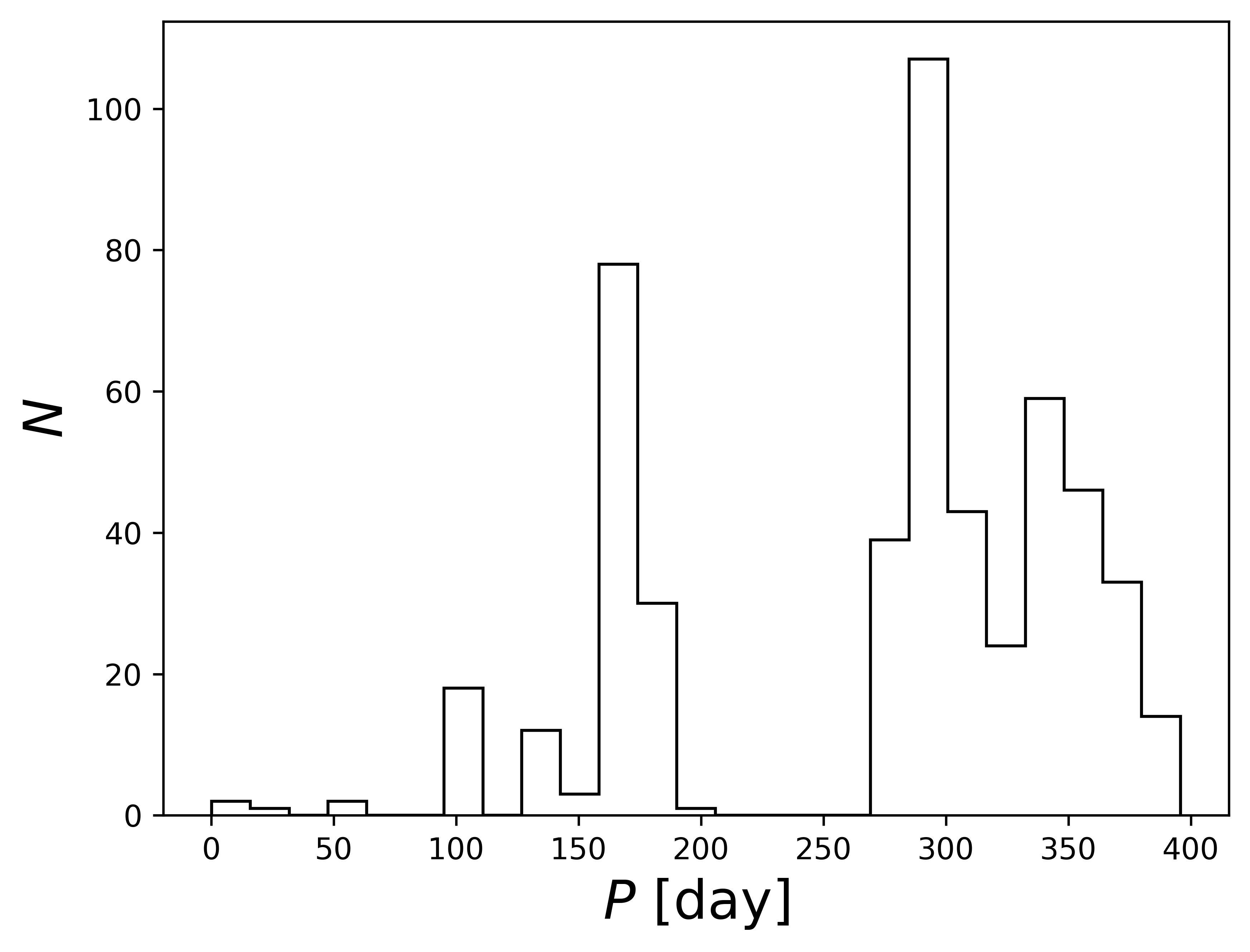}
            \label{fig:2M12pdist}
        }
        \subfloat{
            \includegraphics[width=0.48\textwidth]{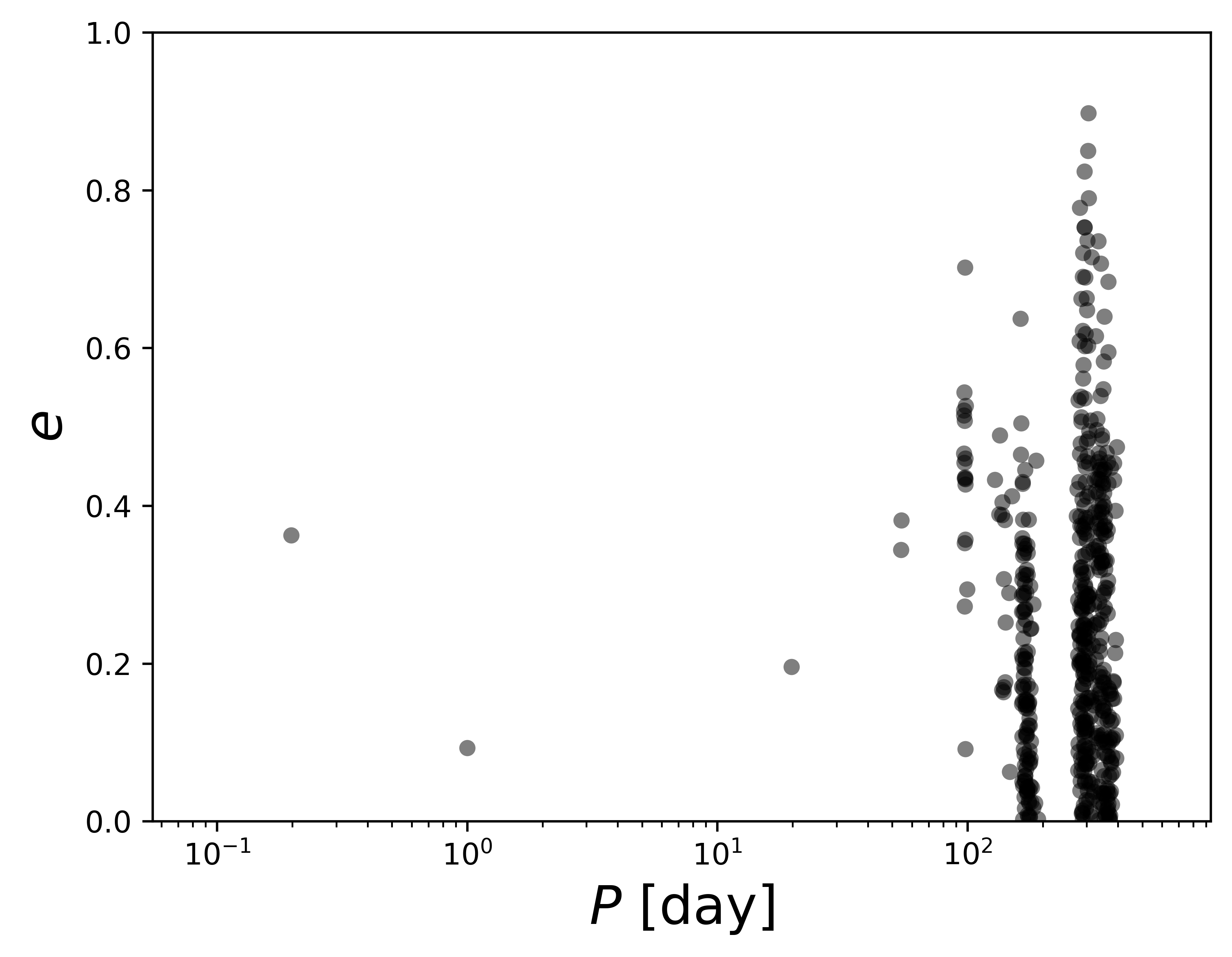}
            \label{fig:2M12pve}
        }

        \caption{Plots used to place limits on the orbital periods of WBs and two PCE systems.  In this example for 2M12423245$-$0646077, two predominant groupings are found in the period versus eccentricity diagram (right) for the 512 possible solutions.  We then place the cautious lower limit of $P = 150$ d as the period distribution (left) clearly shows that a majority of the solutions fall at longer orbital periods.  For some systems, there is more random solutions amongst the groupings.  The limits on the two PCE systems were evaluated with the eccentricity fixed to $e=0$ and with upper rather than lower limits. }
        \label{fig:limit_plots}
\end{figure*}

Additionally, we perform a separate rejection sampling with \textit{The Joker} with a higher than normal number (512 as opposed to 256) requested samples for these systems.  By analyzing the period distribution and a period versus eccentricity diagram for the potential solutions for these systems (examples shown in Figure \ref{fig:limit_plots}), we attempted to find groupings of possible solutions that are indicative of limits to the orbital period of the system.  In the case of WBs, this would be a lower limit based on the smaller RV spread.  This exercise sometimes does not yield conclusive results; therefore, we present lower limits on the orbital period for these WB systems in Column 3 of Table \ref{tab:all_orbits}, but we advise caution in interpreting these results as they may not be indicative of the true solution.  The $\Delta {\rm RV}_{\rm max}$ value presented in Column 2 similarly act as a rough limit of twice the potential velocity semi-amplitude for each system based on the current data.

\subsection{Targets with 2-5 RV Visits}\label{sec:25visits}
There are ten systems that have from two to five high-quality visits in APOGEE.  Two of these systems (2M01575656$-$0244460 and 2M15150334+3628203) have an RV spread of  $\Delta{\rm RV_{\rm max}}>60\,{\rm km\,s}^{-1}$,
which is sufficiently large to  classify them safely as PCE systems, despite the fact that there is insufficient data to derive a full orbital solution.  These two systems were identified as candidate WDMS systems from the SDSS photometric catalog, and we again update the classification to a PCE WDMS candidate; however, 2M01575656$-$0244460 shows a significantly blue GALEX \citep{GALEX2017} FUV-NUV color (-0.26 mag) which is likely indicative of a WD companion.  Additionally, there are two other systems that are well known PCE WDMS systems (2M04322373+1745026 and 2M12154411+5231013; HZ 9 and EG UMa; see \citealt{Rios-Venegas2020} and \citealt{Bleach2000} and references therein, respectively) that were observed.  We do not attempt to add to the previous orbital studies of EG UMa and HZ 9 as we have only have two and three data points, respectively, that are temporally close to one other; however, we do still report the MS star mass derived from our ASPCAP parameters for each.  The mass we derive for EG UMa's MS component is consistent with previous studies, and the mass for HZ 9 is only slightly higher than those shown in \citet{Rios-Venegas2020} and references therein.  Using similar reasoning to that in Section \ref{sec:WBs}, we provisionally classify the remaining six systems as candidate wide-WDMS binaries.  

Additional RV epochs will be necessary for the candidates to confirm these classifications because, for a variety of reasons, the current data are not sufficient to reveal their true nature.  Similar to what was done in Section \ref{sec:WBs}, though, we view period distributions and period eccentricity diagrams in an attempt to place limits on the periods of these systems.  For the WB candidates, we again set cautious lower limits on the period.  For the two PCE candidates, we first fix the eccentricity to $e=0$ (under the same reasoning employed in Sec.~\ref{sec:2M13054173+3037005} and \ref{sec:2M14544500+4626456}) and then place a cautious upper limit to the current data.  As the RV spread on these systems is significantly larger than the WB candidates, this limit is more robust than the lower limits on the WBs; 
however, the number of data points is still small and, therefore, abundant caution should still be exercised in adopting these values until additional data can test the significance of each limit.  

\begin{table*}
    \tiny
    \centering
    
    \caption{Orbital parameters derived using \textit{The Joker} for systems with more than one RV visit.  For WB systems, the period reported is a cautious lower limit on the period.  For two of the marked PCE systems, the period reported corresponds to a cautious upper limit based on the small number of data point; the other marked PCE system has entries that represent one of many possible solutions.  Here $m \sin i$ refers to the WD as the MS star $\left(M_{*}\right)$ is the star being fit by the RV template.  Here $\Delta {\rm RV}_{\rm max}$ serves as a rough limit of twice the velocity semi-amplitude ($K$).  2M22145972$-$0820200 does not have ASPCAP parameters and, therefore, no mass can be derived.}

    \begin{tabular}{@{} lrrrrrrrr @{}}
    \toprule
    
    \multicolumn{1}{c}{APOGEE ID} & 
    \multicolumn{1}{c}{$\rm \Delta RV_{max}$} & 
    \multicolumn{1}{c}{$P$} & 
    \multicolumn{1}{c}{$e$} & 
    \multicolumn{1}{c}{$K$} & 
    \multicolumn{1}{c}{$\gamma$} & 
    \multicolumn{1}{c}{$M_\star$} & 
    \multicolumn{1}{c}{$m \sin i$} &  
    \multicolumn{1}{c}{$T_{\rm eff, WD}\!^{\dagger}$} \\
    
    \multicolumn{1}{c}{} & 
    \multicolumn{1}{c}{\tiny{$\left[\rm km\,s^{-1}\right]$}} & 
    \multicolumn{1}{c}{\tiny{$\rm\left[days\right]$}} & 
    \multicolumn{1}{c}{} & 
    \multicolumn{1}{c}{\tiny{$\left[\rm km\,s^{-1}\right]$}} & 
    \multicolumn{1}{c}{\tiny{$\left[\rm km\,s^{-1}\right]$}} & 
    \multicolumn{1}{c}{\tiny{$\left[M_{\odot}\right]$}} & 
    \multicolumn{1}{c}{\tiny{$\left[M_{\odot}\right]$}} &
    \multicolumn{1}{c}{[K]} \\

    \midrule
    \multicolumn{9}{c}{\tiny{PCE Systems}} \\
    \midrule
    2M01575656$-$0244460$^{a,e}$ & $75.13\pm0.07$ & 3.5 & 0 & - & - & $0.534\pm0.034$ & - & -\\
    2M04322373$+$1745026 & - & - & - & - & - & $0.440\pm0.028$ & - & 29727$\pm$691\\
    2M10243847$+$1624582 & $293.19\pm11.63$ & $0.5258733\pm0.0000056$ & $0.0295660\pm0.0143324$ & $145.6\pm2.3$ & $-8.5\pm1.7$ & $0.423\pm0.027$ & $0.537\pm0.013$ & 15246$\pm$482\\
    2M10552625$+$4729228 & $148.07\pm0.08$ & $2.1866303\pm0.0023683$ & $0.0033047\pm0.0157235$ & $82.4\pm1.0$ & $8.2\pm0.9$ & $0.446\pm0.029$ & $0.476\pm0.009$ & 26801$\pm$113\\
    2M11463394$+$0055104 & $373.00\pm2.07$ &  $0.4087104\pm0.0000114$ &  $0.0272674\pm0.0135032$ & $186.8\pm2.1$ & $-12.5\pm2.8$ & $0.440\pm0.028$ & $0.717\pm0.014$ & -\\
    2M12154411$+$5231013 & - & - & - & - & - & $0.485\pm0.031$ & - & 15601$\pm$586\\
    2M13054173$+$3037005$^{b,e}$ & $68.26\pm0.15$ & $0.2165179\pm0.0000015$ & 0 & $48.2\pm0.4$ & $92.0\pm0.2$ & $0.453\pm0.029$ & $0.091\pm0.001$ & -\\
    2M14544500$+$4626456$^{e}$ & $99.53\pm0.04$ & $15.0957084\pm0.0000698$ & 0 & $53.41\pm0.06$ & $-22.20\pm0.05$ & $0.488\pm0.031$ & $0.693\pm0.001$ & -\\
    2M15150334$+$3628203$^{a,e}$ & $228.08\pm0.11$ & 13.5 & 0 & - & - & $0.489\pm0.031$ & - & -\\
    
    \midrule
    \multicolumn{9}{c}{\tiny{WB Systems}} \\
    \midrule
    2M03160020+0009462$^{c,e}$ & $0.87\pm0.02$ & - & - & - & - & $0.483\pm0.031$ & - & 19416$\pm$1144\\
    2M03452349$+$2451029 & $1.08\pm0.01$ & 20.0 & - & - & - & $0.502\pm0.032$ & - & -\\
    2M08094855$+$3221223 & $0.91\pm0.05$ & 200.0 & - & - & - & $0.436\pm0.028$ & - & -\\
    2M09463250$+$3903015$^{c}$ & $0.37\pm0.08$ & - & - & - & - & $0.486\pm0.031$ & - & -\\
    2M12423245$-$0646077$^{e}$ & $2.58\pm0.07$ & 150.0 & - & - & - & $0.478\pm0.031$ & - & -\\
    2M13115337$+$1549147 & $1.01\pm0.12$ & 100.0 & - & - & - & $0.436\pm0.028$ & - & -\\
    2M14244053$+$4929580 & $3.52\pm0.59$ & 100.0 & - & - & - & $0.438\pm0.028$ & - & 36572$\pm$475\\
    2M14551261$+$3810342$^{d,e}$ & $0.43\pm0.01$ & - & - & - & - & $0.431\pm0.028$ & - & 14728$\pm$2041\\
    2M15041191$+$3658150 & $1.03\pm0.06$ & 100.0 & - & - & - & $0.440\pm0.028$ & - & -\\
    2M18454771$+$4431148$^{e}$ & $0.43\pm0.00$ & 400.0 & - & - & - & $0.913\pm0.058$ & - & 33740$\pm$3576\\
    2M22145972$-$0820200$^{c,e}$ & $0.04\pm 0.01$ & - & - & - & - & - & - & -\\
    2M22200576$-$0418445$^{c,e}$ & $0.11\pm0.00$ & - & - & - & - & $0.442\pm0.028$ & - & -\\
    
    \bottomrule
    
    \multicolumn{8}{@{}l}{\textbf{\scriptsize{Notes}}}\\
    \multicolumn{8}{@{}l}{\tiny{$^{\dagger}$ Estimate from the system's respective catalog}} \\
    \multicolumn{8}{@{}l}{\tiny{$^{a}$ Here $P$ refers to a cautious upper limit with an eccentricity fixed to $e=0$}} \\
    \multicolumn{8}{@{}l}{\tiny{$^{b}$ Here the entries are for one of many possible solutions}} \\
    \multicolumn{8}{@{}l}{\tiny{$^{c}$ Here there are only two RVs that are too close temporally to place accurate limits}} \\
    \multicolumn{8}{@{}l}{\tiny{$^{d}$ Here the data produced incommensurate potential solutions such that no limit can be placed on the period}} \\
    \multicolumn{8}{@{}l}{\tiny{$^{e}$ Candidate system}}
    
    \end{tabular}
    
    \label{tab:all_orbits}
\end{table*}

\subsection{Targets with a Single Visit}\label{sec:onevisit}
There are seven of the remaining 28 systems that only received one high-quality visit throughout the course of the APOGEE survey and two that were targeted by APOGEE, but for which the data were insufficient for either the derivation of a radial velocity and/or ASPCAP parameters.
No RV variations for these systems can be determined in this work, and it is unlikely that these systems will receive additional APOGEE visits in the future; however, the RV measurement (when present) for each system is included in Appendix \ref{app:RVs} for completeness.  In the future, these measurements can be combined with those from dedicated follow-up studies or other spectroscopic sky surveys.
Available APOGEE measurements for these systems are listed in Table \ref{tab:onevisit} in Appendix \ref{app:RVs}.

\section{Metallicity Distribution of WDMS Systems} \label{sec:MDF}

The APOGEE database, featuring chemical abundances derived from high resolution spectroscopy and multi-epoch radial velocities, provides a unique opportunity to explore correlations between stellar chemistry and binary star architectures \citep[e.g.,][]{Mazzola2020}.  The metallicities of M dwarfs, which constitute most of the MS companions in the present sample, are notoriously difficult to measure \citep[e.g.,][]{Newton2014}, however \citet{Souto2020} show that ASPCAP is sufficient to measure metallicities for these low-mass stars within about 0.1 -- 0.2 dex.  The ASPCAP fits do not take into account irradiation effects from the WD or tidal distortion effects in determining stellar parameters, but, based on available temperature estimates for the WDs in our sample, we do not expect these to significantly affect the metallicities of the M dwarfs in our sample.  We therefore exploit these ASPCAP parameters here to explore the metallicity distribution of our vetted and cleaned WDMS sample, which is a parameter space not often explored for these systems, but one that has been questioned as potentially correlated to other system characteristics.  For example, among the few studies looking at WDMS metallicities, \citet{Rebassa-Mansergas2018} explored the age-metallicity relation for a dataset of 23 WDMS systems and found no significant correlation between the MS star's [Fe/H] value and the WD's age. Meanwhile, \citet{Parsons2018} measured metallicities for thirteen eclipsing, PCE WDMS systems and did not find clear evidence for the metallicity to be the cause of over-inflation in the radii of their M dwarf sample.

\begin{figure}
    \centering
    \includegraphics[width=\linewidth]{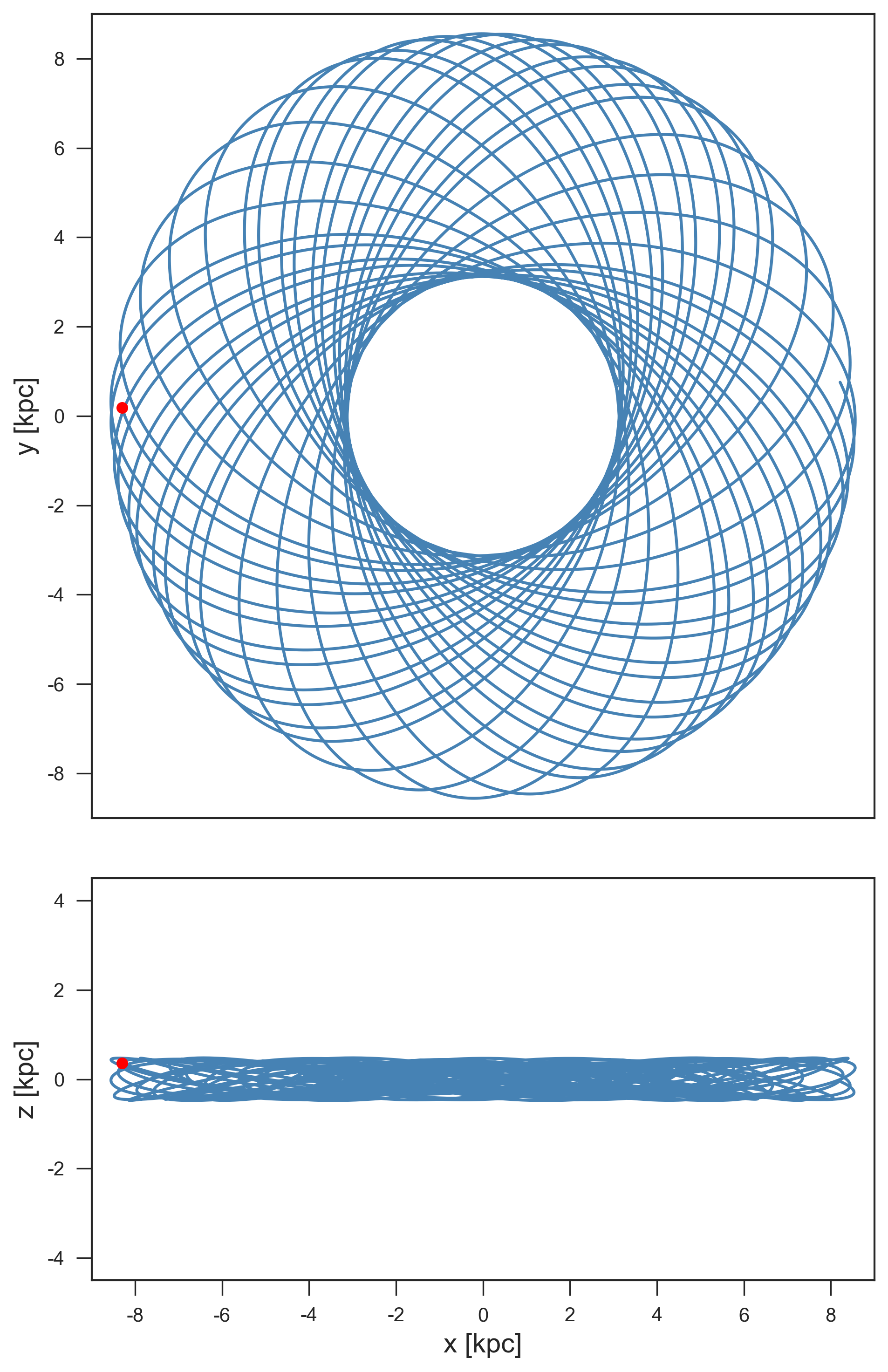}
    \caption{The calculated orbit for 2M14244053+4929580 aged 5 Gyr backwards from the present day, shown in the  Galactic Cartesian coordinate system with a 1:1 aspect ratio in both projections to emphasize the planar nature of the orbit.  The red dot marks the system's current location.  It is clear that this system has a very disk-like orbit despite being having a metallicity ([Fe/H] $\sim$ -1.4) rather typical of the Galactic halo.}
    \label{fig:sd_orbit}
\end{figure}

\begin{figure}[h]
    \centering
    \includegraphics[width=\linewidth]{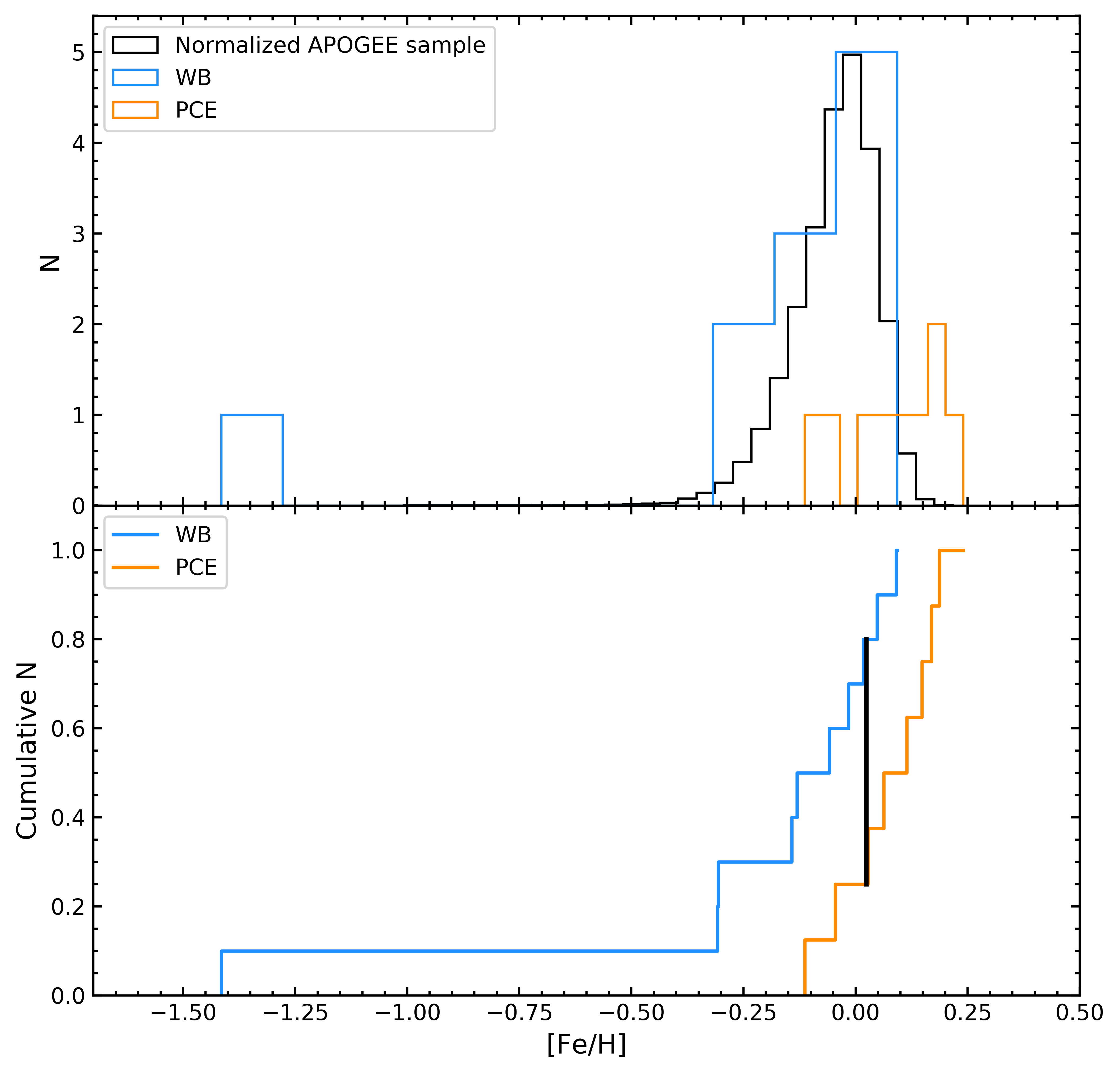}
    \caption{\textit{Top:} The metallicity distribution of the WB and PCE systems listed in Table \ref{tab:all_orbits} shown in blue and orange, respectively.  For comparison we also show in black the distribution of APOGEE MS stars sharing similar effective temperatures and surface gravities as the MS primaries in the WB and PCE samples. \textit{Bottom:}  Cumulative distribution for the WB and PCE in blue and orange, respectively.
    The thick, black, vertical line shows the KS $D$-statistic, the maximum distance between the two distributions.
    }
    \label{fig:MDF}
\end{figure}

Figure \ref{fig:MDF} shows the metallicity distribution of our 21 systems, separated into their respective classifications of WB or PCE systems.  As discussed in Section \ref{sec:cleaning}, there is one system (2M14244053+4929580), found here to be a WB, that is extremely metal-poor ([Fe/H] $= -1.4$) compared to all other systems in this paper.  In fact, this [Fe/H] is $\sim$0.6 dex more metal poor than that for the most metal-poor system from \citet{Rebassa-Mansergas2018}.  Recently, however, \citet{Rebassa-Mansergas2019NatAs} studied another eclipsing, PCE WDMS system, SDSS J235524.29+044855.7, a short period binary containing a halo subdwarf with [Fe/H] = $-1.55\pm0.25$, which then relegates 2M14244053+4929580 to the second most metal-poor WDMS system reported to date.  As in the case of the former system, a metallicity this low is typically indicative of a star system that belongs to the Milky Way halo population; however, a kinematical analysis does not support that assumption.  We used the \texttt{astro-gala} \citep{Gala} Python package along with the combined APOGEE RV and \textit{Gaia} parallax and proper motion to calculate the system's orbit.  The result, shown in Figure \ref{fig:sd_orbit}, reveals the system to have a rather prototypical (old) disk star orbit, with a maximum excursion from the disk of only 0.48 kpc over a radial variation ranging from 3 to 10 kpc.  The juxtaposition of this rather planar orbit with such a low metallicity makes 2M14244053+4929580 a somewhat unusual system (even ignoring that it is also a WDMS binary).  Because it seems an outlier, in the following analyses we consider statistics that both include and exclude this unusual system.

At the other end of the MDF, four of the PCE systems, and many WBs, appear to have super-solar [Fe/H] values.
However, the overall metallicity distribution of PCE systems seems to be significantly higher than that for WB systems.  This is born out by the medians and dispersions of the two groups, which are $0.114$ dex with $\sigma = 0.109$ dex for the PCE, but $-0.059$ dex with $\sigma = 0.409$ dex (-0.037 dex and $\sigma = 0.141$ dex excluding the subdwarf) for the WBs.
The latter group have a metallicity distribution function similar to that for the full APOGEE sample of MS stars with similar effective temperatures and surface gravities to those of the WDMS sample (shown for comparison in Figure \ref{fig:MDF}).
These values demonstrate that the median metallicities of the two systems are separated by $\sim 1\sigma$.

A Kolomogorov-Smirnov (KS) test\footnote{All statistical tests here  make use of the\\ \texttt{scipy.stats} \citep{SciPy-NMeth} Python module.}
of the two distributions with the hypothesis that the WB and PCE systems derive from the same parent distribution yields a KS statistic of $D=0.556$ (see Fig. \ref{fig:MDF}, bottom panel) and a $p$-value of $p=0.058$, meaning the hypothesis can be rejected at the 90\% confidence level.
Because KS tests are not always sensitive enough to determine whether two distributions are independent, we also perform an Anderson-Darling (AD) test, which is more sensitive to a distribution's wings and yields a standardized test statistic of $T=3.598$.  This value allows us to firmly reject our hypothesis at the 97.5\% confidence level. 
Repeating these tests with the metal-poor WB removed does not change the KS $D$ statistic, and, with only a slight change in sample size,
yields similar results:  
a  KS $p$-value of $0.084$, implying a rejection of the null hypothesis at the 90\% confidence level, and an AD test with $T=2.988$, rejecting the null hypothesis firmly at the 97.5\% confidence level .
Meanwhile, in contrast, a KS test 
comparing the WBs and the APOGEE MS star sample show them to be essentially indistinguishable.  

\begin{figure}[h]
    \centering
    \includegraphics[width=\linewidth]{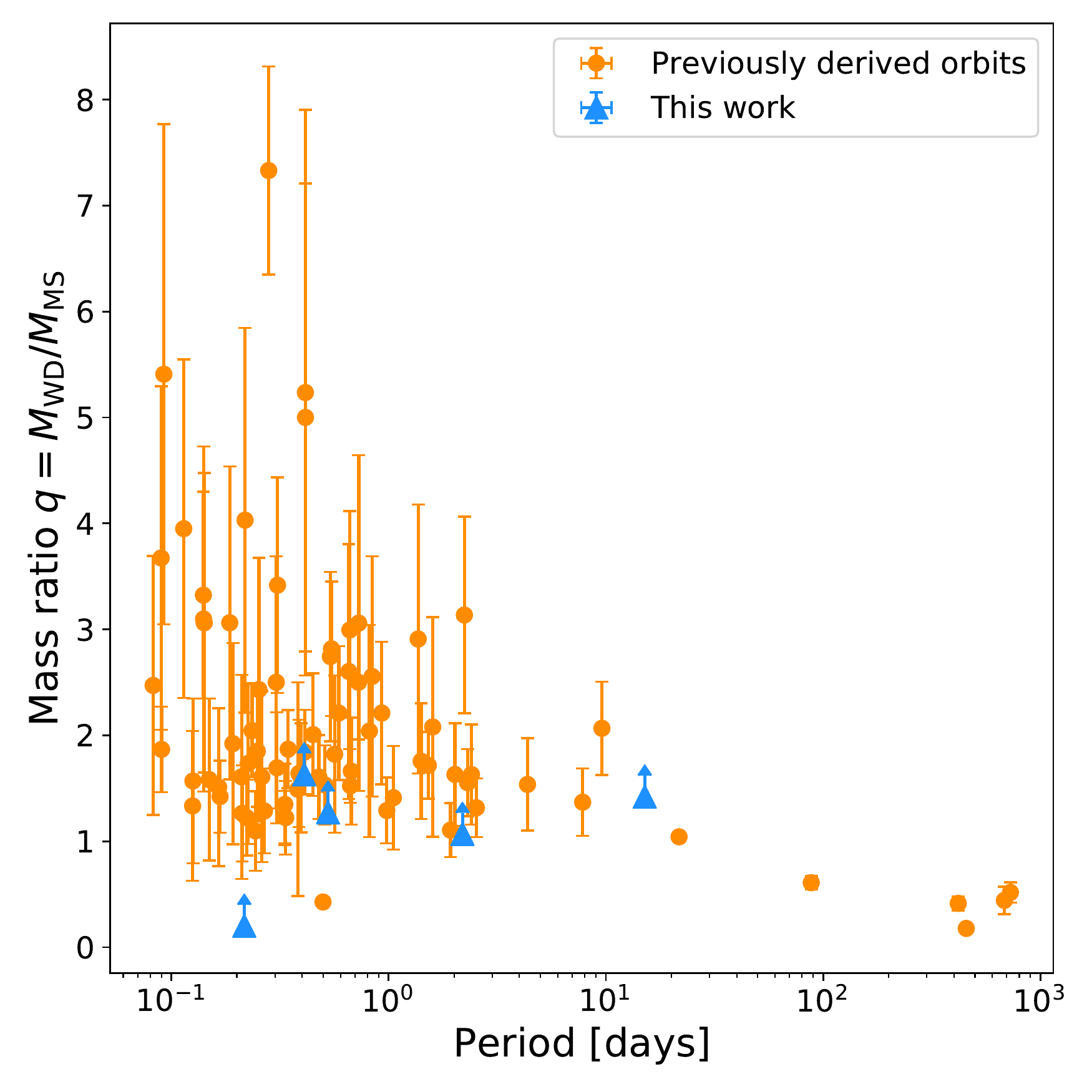}
    \caption{
    Mass ratio, $q$, versus period for PCE systems with previously derived orbital parameters and component masses (orange). The five systems with orbital parameters derived in this work are also shown (blue). The mass ratios for the latter 
    represent lower limits on $q$ because we derive the minimum WD mass from the Keplerian orbital parameters.}

    \label{fig:MvP}
\end{figure}

The difference in the [Fe/H] distribution of WB and PCE systems could speculatively point to some sort of alteration of a system's surface chemistry during the CE phase.  However, \citet{Hjellming1991} showed that CEs have much higher specific entropies than the surface of the secondary star, meaning the companion should be thermally isolated from the CE and, as a result, almost no accretion takes place.  Thus, it would seem that the M dwarfs would not be polluted by metals during the CE phase.  The previously discussed results of \citet{Parsons2018} that M dwarf radii in PCE systems were indistinguishable from the radii of field M dwarfs would also support this hypothesis; however, it is worth noting that some M dwarfs in PCE systems in the center of planetary nebulas are substantially inflated \citep[e.g.,][]{Jones2015}.  This may be evidence that the companions are actually slightly altered by the CE phase, but only affected for a short while.  Could a possible explanation for the obvserved metallicity difference be that the effects of close binarity alter the metallicities?  Could rapid rotation in close systems affect the metallicities? Of course, the results here must be considered tentative given the overall small WDMS sample, especially given selection biases in the parent samples (see Section \ref{sec:cleaning}).  Nevertheless, the metallicity differences seen here between the PCE and WB groups offer a tantalizing incentive for further studies of the chemistry of WDMS systems.  A significant contribution in this direction can be expected from the APOGEE survey itself, where a large number of newly discovered WDMS candidates have been found \citep[][Anguiano et al., in prep]{Anguiano2020}.  Moreover, not only do the majority of these systems have well characterized metallicities by APOGEE, but more detailed chemical abundance patterns as well, a unique opportunity for WDMS surveys to gain key insights into the role of chemical composition in the evolution of WDMS systems.

\section{Summary} \label{sec:summary}
We have presented an analysis of the 45 candidate or confirmed WDMS systems identified by SDSS and LAMOST that also lie in the APOGEE survey.
The results of our investigation of these systems are as follows:

\begin{itemize}
\item By examining the APOGEE-derived stellar parameters for the putative MS star in each system (Fig.~\ref{fig:CMD}), APOGEE identifies three to be RG contaminants and 14 to be YSO contaminants of the 45 stars in the parent sample (Sec.~\ref{sec:cleaning}).  We propose various reasons why YSOs may have a high contamination rate within photometrically-selected WDMS catalogs and have thereby come to constitute almost a third of our starting sample.

\item After imposing quality cuts on acceptable APOGEE RV measurements, we used \textit{The Joker} to derive or place limits on the orbital parameters for 14 of the WDMS systems having more than two visits (Sec. \ref{sec:RVs}). In addition, using the \citet{Torres2010} relations and the APOGEE stellar parameters, we derive the MS star mass for each system, when that is possible (Sec.~\ref{sec:stellar_masses}). 

\item 
A key result of our orbital analysis is the confirmation of nine previously confirmed or candidate PCE systems: two that are well known (Sec.~\ref{sec:25visits}), three that are newly confirmed (Sec.~\ref{sec:2M10243847+1624582}-\ref{sec:2M11463394+0055104}), and four that are newly discovered (Sec.~\ref{sec:2M13054173+3037005}-\ref{sec:2M14544500+4626456}, \ref{sec:25visits}).  For three of these systems we present robust orbital solutions (Sec.~\ref{sec:2M10243847+1624582}-\ref{sec:2M11463394+0055104}), while a reasonable solution is given for another (Sec.~\ref{sec:2M14544500+4626456}) and a lower-limit solution for the remaining one (Sec.~\ref{sec:2M13054173+3037005}), thereby adding to the relatively small \citep[$\sim$90; e.g.,][]{Nebot2011,Rebassa-Mansergas2016,Parsons2015} number of PCE systems having derived orbital parameters\footnote{Orbital parameters and component masses for 90 PCE systems are available at \url{https://www.sdss-wdms.org/}.} and the $\sim$120 having at least spectroscopically or photometrically defined orbital periods \citep[e.g.,][]{Ren2018}.

\item
Though the mass ratios derived in this work represent lower-limits for $q$, they tend to imply low mass ratios (with $1 \lesssim q \lesssim 2$), which is similar to what has been found for a majority of previous solutions for other systems. 

\item 
While the three systems for which we derive robust solutions have orbital periods typical of most PCE systems, our solution for 2M14544500+4626456, if confirmed to be a WDMS, would make it a PCE WDMS binary with the second longest period known for typical, compact systems (Sec.~\ref{sec:2M14544500+4626456}).  We also have contributed 12 tentative WB classifications (Sec.~\ref{sec:WBs}), however, these may, of course, change with additional RV data in the future.

\item
We report 2M14244053+4929580 to be, by far, the most metal-poor WDMS system known to date, with an [Fe/H] near the mean for the Galactic halo, of which we initially suspected it to be a member (Sec.~\ref{sec:cleaning}, \ref{sec:MDF}).  However, an analysis of the orbit of this binary (Fig.~\ref{fig:sd_orbit}) shows it to have an orbit more characteristic of a Galactic disk star (Sec.~\ref{sec:MDF}). 

\item
The WB stars in our sample have an MDF that is significantly skewed to lower metallicities than the PCE stars (Fig. \ref{fig:MDF}). We speculate on reasons for this MDF difference, but also caution that the analysis is based on a relatively small sample (Sec. \ref{sec:MDF}).  

\end{itemize}

The results of the present exploration of previously known WDMS systems demonstrate the efficacy of APOGEE data for not only characterizing the orbital properties of such systems, but also for identifying new WDMS candidates by their RV variability, in particular, those sources with short period solutions that can add to the known number of PCE WDMS binaries having derived orbital parameters.  In a future paper (Anguiano et al., in prep.) we will not only demonstrate this by further exploiting the APOGEE catalog, but, in so doing, substantially increase the number of known and characterized WDMS systems.

\acknowledgments

We thank the anonymous referee for insightful comments that significantly improved the content of the manuscript.  KAC, HML, BA, SRM, and DJM acknowledge support from National Science Foundation grant AST-1616636.  BA also acknowledges the AAS Chr{\'e}tien International Research  Grant. ND would like to acknowledge that this material is based upon work supported by the National Science Foundation under Grant No. 1616684.  ARL acknowledges partial financial support for APOGEE2 provided in Chile by Comisi{\'o}n Nacional de Investigaci{\'o}n Cient{\'\i}fica y Tecnol{\'o}gica (CONICYT) through the FONDECYT project 1170476 and QUIMAL Project 130001.

Funding for the Sloan Digital Sky Survey IV has been provided  by  the  Alfred  P.  Sloan  Foundation,  the  U.S.Department of Energy Office of Science, and the Participating  Institutions.   SDSS  acknowledges  support  and resources from the Center for High-Performance Computing at the University of Utah.  The SDSS web site is www.sdss.org.

SDSS is managed by the Astrophysical Research Consortium for the Participating Institutions of the SDSS Collaboration including the Brazilian Participation Group, the Carnegie Institution for Science, Carnegie Mellon University, the Chilean Participation Group, the French Participation Group, Harvard-Smithsonian Center for Astrophysics, Instituto de Astrofísica de Canarias, The Johns Hopkins University, Kavli Institute for the Physics and Mathematics of the Universe (IPMU) / University of Tokyo, the Korean Participation Group, Lawrence Berkeley National Laboratory, Leibniz Institut für Astrophysik Potsdam (AIP), Max-Planck-Institut für Astronomie (MPIA Heidelberg), Max-Planck-Institut für Astrophysik (MPA Garching), Max-Planck-Institut für Extraterrestrische Physik (MPE), National Astronomical Observatories of China, New Mexico State University, New York University, University of Notre Dame, Observatório Nacional / MCTI, The Ohio State University, Pennsylvania State University, Shanghai Astronomical Observatory, United Kingdom Participation Group, Universidad Nacional Autónoma de México, University of Arizona, University of Colorado Boulder, University of Oxford, University of Portsmouth, University of Utah, University of Virginia, University of Washington, University of Wisconsin, Vanderbilt University, and Yale University.

This work presents results from the European Space Agency (ESA) space mission \textit{Gaia}. \textit{Gaia} data are being processed by the \textit{Gaia} Data Processing and Analysis Consortium (DPAC). Funding for the DPAC is provided by national institutions, in particular the institutions participating in the \textit{Gaia} MultiLateral Agreement (MLA). The \textit{Gaia} mission website is https://www.cosmos.esa.int/gaia. The \textit{Gaia} archive website is https://archives.esac.esa.int/gaia.

This work made use of the SIMBAD database, operated at CDS, Strasbourg, France; the VizieR catalog access tool, CDS, Strasbourg, France.

This work made use of TOPCAT, an interactive graphical viewer and
editor for tabular data \citep{TOPCAT}.

%



\software{APOGEE reduction software \citep{Nidever2015AJ}, 
        ASPCAP \citep{Garcia-Perez2016AJ,Jonsson2020}, 
        The Joker \citep{Price-Whelan2017ApJ,Price-Whelan2020ApJ}, 
        TOPCAT \citep{TOPCAT}, 
        astro-gala \citep{Gala}
        }



\newpage
\bibliography{references}{}
\bibliographystyle{aasjournal}

\newpage
\appendix
\section{Systems with One or less High--Quality Visits}\label{app:RVs}

The seven systems shown in Table \ref{tab:onevisit} were targeted by APOGEE, but, in the end, received zero or one quality visit over the duration of the survey. These systems are not likely to receive additional visits in the APOGEE-2 survey, but obviously would benefit from
additional data.  The table summarizes what is known about these systems from the current APOGEE data in hand.

\begin{table*}[h]
    \scriptsize
    \centering
    
    \caption{Data for APOGEE systems in the cleaned sample with zero or one quality visit.
    }

    \begin{tabular}{@{} lrrrr @{}}
    \toprule
    
    \multicolumn{1}{c}{Ref. ID} & 
    \multicolumn{1}{c}{JD} & 
    \multicolumn{1}{c}{RV} &
    \multicolumn{1}{c}{$M_{*}\!^{a}$} \\
    
    \multicolumn{1}{c}{} & 
    \multicolumn{1}{c}{} & 
    \multicolumn{1}{c}{\tiny{[$\rm km\,s^{-1}$]}} &
    \multicolumn{1}{c}{\tiny{[$M_{\odot}$]}} & 
    \multicolumn{1}{c}{} \\

    \midrule
    2M08424235$+$5128575 & 2457046.86882 & 38.93282 $\pm$ 18.471933 &  & \\
    2M08531787$+$1147595 & 2458183.56201 & 80.99747 $\pm$ 0.09029441 & $0.463\pm0.030$ & \\
    2M11241545$+$4558412 & 2457151.67308 & 8.04085 $\pm$ 0.0831778 & $0.478\pm0.031$ & \\
    2M12333939$+$1359439 & 2458617.65376 & 85.52377 $\pm$ 1.351955 & & \\
    2M13090450$+$1411351$^{b}$ & & & & \\
    2M13463968$-$0031549$^{b}$ & & & & \\
    2M15104562$+$4048271 & 2457898.68379 & -22.661102 $\pm$ 0.1061763 & $0.413\pm0.026$ & \\

    \bottomrule
    
    \multicolumn{5}{@{}l}{\textbf{\scriptsize{Notes}}}\\
    \multicolumn{5}{@{}l}{\tiny{$^{a}$ If MS star with ASPCAP parameters available}} \\
    \multicolumn{5}{@{}l}{\tiny{$^{b}$ Received no high-quality visits}}
    
    \end{tabular}
    
    \label{tab:onevisit}
\end{table*}

\end{document}